\documentclass[10pt,twocolumn,letterpaper]{article}

\usepackage{iccv}
\usepackage{times}
\usepackage{epsfig}
\usepackage{amsmath}
\usepackage{amssymb}
\usepackage{mynotation}
\usepackage{xcolor}
\usepackage{booktabs}

\usepackage{comment}
\usepackage{bbm}
\usepackage{mathtools}
\usepackage{algorithm}
\usepackage[noend]{algpseudocode}
\usepackage[caption=false]{subfig}
\usepackage{graphicx}

\usepackage[pagebackref=true,breaklinks=true,letterpaper=true,colorlinks,bookmarks=false]{hyperref}

\iccvfinalcopy 

\ificcvfinal\fi

\usepackage{capt-of,etoolbox}
\makeatletter
\apptocmd\@maketitle{{\introfig{}\par}}{}{}
\makeatother

\begin{document}

\title{Deep Reparametrization of Multi-Frame Super-Resolution and Denoising}

\author{Goutam Bhat \qquad Martin Danelljan \qquad Fisher Yu \qquad Luc Van Gool \qquad Radu Timofte\\
	Computer Vision Lab, ETH Zurich, Switzerland\\
}

\newcommand{\introfig}{
\centering\vspace{-4mm}%
	\newcommand{\wid}{0.99\textwidth}%
	\includegraphics*[trim = 0 0 0 2, width = \wid]{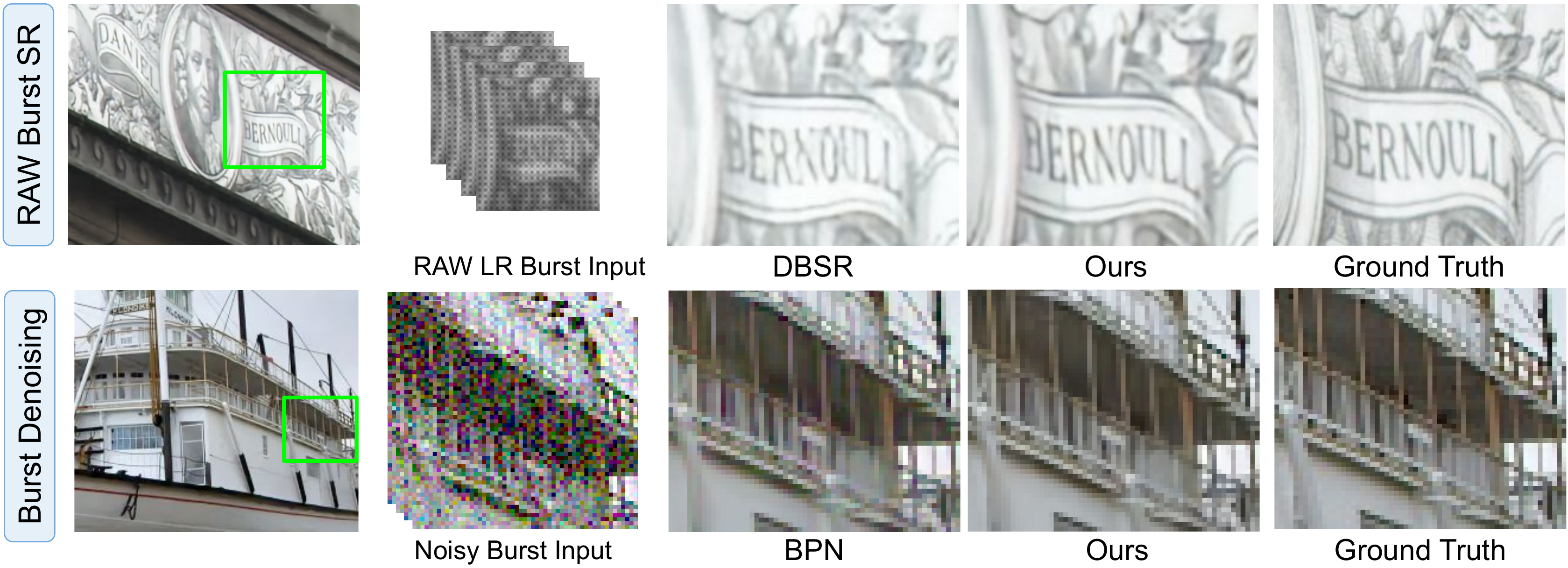}\hspace{1mm}
	\vspace{-5mm}
	\captionof{figure}{We propose a deep reparametrization of the classical MAP objective \eqref{eq:map_intro} for multi-frame image restoration. Our general formulation minimizes a learned reconstruction error in a deep latent space. The proposed approach outperforms previous state-of-the-art methods DBSR~\cite{Bhat2021DeepBS} and BPN~\cite{Xia2020BasisPN} on the RAW burst super-resolution (top) and burst denoising (bottom) tasks, respectively.}%
	\label{fig:intro}%
\vspace{6mm}}

\maketitle
\ificcvfinal\thispagestyle{empty}\fi

\begin{abstract}
We propose a deep reparametrization of the maximum a posteriori formulation commonly employed in multi-frame image restoration tasks.
Our approach is derived by introducing a learned error metric and a latent representation of the target image, which transforms the MAP objective to a deep feature space. The deep reparametrization allows us to directly model the image formation process in the latent space, and to integrate learned image priors into the prediction.
Our approach thereby leverages the advantages of deep learning, while also benefiting from the principled multi-frame fusion provided by the classical MAP formulation. 
We validate our approach through comprehensive experiments on burst denoising and burst super-resolution datasets.
Our approach sets a new state-of-the-art for both tasks, demonstrating the generality and effectiveness of the proposed formulation.\vspace{-1mm}
\end{abstract}

\newcommand{\parsection}[1]{\vspace{0.5mm}\noindent\textbf{#1:}~}
\newcommand{\pred}{\hat{y}}
\newcommand{\weightpred}{W}
\newcommand{\predenc}{z}
\newcommand{\yc}{y}
\newcommand{\emb}{e}
\newcommand{\shift}{\phi}
\newcommand{\shiftd}{\Bar{\shift}}
\newcommand{\reg}{\mathcal{R}}
\newcommand{\regenc}{\mathcal{Q}}
\newcommand{\h}{H}
\newcommand{\he}{\hat{\h}}
\newcommand{\motion}{m}
\newcommand{\motiond}{\Bar{\motion}}
\newcommand{\solver}{A}
\newcommand{\enc}{E}
\newcommand{\dec}{D}
\newcommand{\hd}{\Bar{\h}}
\newcommand{\noise}{\eta}
\newcommand{\noisemapped}{\chi}
\newcommand{\noisedist}{p_\eta}
\newcommand{\modelinit}{P}
\newcommand{\first}[1]{\textbf{\textcolor{red}{#1}}}
\newcommand{\second}[1]{\textbf{\textcolor{blue}{#1}}}


\section{Introduction}
Multi-frame image restoration (MFIR) is a fundamental computer vision problem with a wide range of important applications, including burst photography~\cite{Bhat2021DeepBS,Hasinoff2016BurstPF,Liba2019HandheldMP,Wronski2019HandheldMS} and remote sensing~\cite{Deudon2020HighResnetRF,Kawulok2019DeepLF,Molini2020DeepSUMDN}. Given multiple degraded and noisy images of a scene, MFIR aims to reconstruct a clean, sharp, and often higher-resolution output image. By effectively leveraging the information contained in different input images, MFIR approaches are able to reconstruct richer details that cannot be recovered from a single image.

As a widely embraced paradigm~\cite{Elad1997RestorationOA,Farsiu2004MultiframeDA,Kokkinos2019IterativeRC,Peleg1987ImprovingIR}, MFIR is addressed by first modelling the image formation process as, \mbox{$x_i = \h\big( \shift_{\motion_i}(\yc )\big) + \noise_i$}. 
In this model, the original image $y$ is affected by the scene motion $\shift_{\motion_i}$, image degradation $\h$, and noise $\noise_i$, resulting in the observed image $x_i$.
Assuming the noise $\noise_i$ follows an i.i.d.\ Gaussian distribution, the original image $\yc$ is reconstructed from the set of noisy observations $\{x_i\}_1^N$ by finding the maximum a posteriori (MAP) estimate,
\begin{equation}
\label{eq:map_intro}
    \pred = \argmin_{y} \sum_{i=1}^{N} \big\|x_i - \h\left(\shift_{\motion_i}(y)\right) \big\|_2^2 + \reg(y) \,,
\end{equation}
where $\reg(y)$ is the imposed prior regularization.


While the MAP formulation \eqref{eq:map_intro} has enjoyed much popularity, there are several challenges when employing it in real-world settings. The formulation \eqref{eq:map_intro} assumes that the degradation operator $\h$ is known, which is not often the case. Moreover, it requires manually tuning the regularizer $\reg(y)$ for good performance~\cite{Farsiu2004MultiframeDA,Farsiu2004FastAR,Gotoh2004DirectSA}. 
Despite these shortcomings, the MAP formulation \eqref{eq:map_intro} provides an elegant modelling of the MFIR problem, and a principled way of fusing information from multiple frames. 
This inspires us to formulate a deep MFIR method that leverages the compelling advantages of \eqref{eq:map_intro}, while also benefiting from the end-to-end learning of the degradation operator $\h$ and the regularizer $\reg$. 

We propose a deep reparametrization of the classical MAP objective \eqref{eq:map_intro}.
Our approach is derived as a generalisation of the image space reconstruction problem \eqref{eq:map_intro}, by transforming the MAP objective to a deep feature space.
This is achieved by first introducing an encoder network that replaces the $L_2$ norm in \eqref{eq:map_intro} with a learnable error metric, providing greater flexibility. We then reparametrize the target image $y$ with a decoder network, allowing us to solve the optimization problem in a learned latent space. The decoder integrates strong learned image priors into the prediction, effectively removing the need of a manually designed regularizer $\reg$. Our deep reparametrization also allows us to directly learn the effects of complex degradation operator $\h$ in the deep latent space of our formulation. To further improve the robustness of our model to \eg varying noise levels and alignment errors, we introduce a network component that estimates the certainty weights of all observations in the objective.

We validate the proposed approach through extensive experiments on two multi-frame image restoration tasks, namely RAW burst super-resolution, and burst denoising. Our approach sets a new state-of-the-art on both tasks by outperforming recent deep learning based approaches (see Fig.~\ref{fig:intro}). We further perform extensive ablative experiments, carefully analysing the impact of each of our contributions.

\section{Related Work}
\parsection{Multi-Frame Super-Resolution} MFSR is a well-studied problem, with more than three decades of active research. Tsai and Huang~\cite{Tsai1984MultiframeIR} were the first to propose a frequency-domain based solution for MFSR. Peleg~\etal~\cite{Peleg1987ImprovingIR} and Irani and Peleg~\cite{Irani1991ImprovingRB} proposed an iterative approach based on an image formation model. Here, an initial guess of the SR image is obtained and then refined by minimizing a reconstruction error. Several works~\cite{Bascle1996MotionDA,Elad1997RestorationOA,Hardie1998HighResolutionIR,Schultz1996ExtractionOH} extended the objective in~\cite{Irani1991ImprovingRB} with a regularization term to obtain a maximum a posteriori (MAP) estimate of the HR image. Robustness to outliers or varying noise levels were further addressed in~\cite{Farsiu2004MultiframeDA,Zomet2001RobustS}. 
 
The  aforementioned approaches assume that the image formation model, as well as the motion between input frames can be reliably estimated. Several works address this limitation by jointly estimating these unknown parameters~\cite{Faramarzi2013UnifiedBM,He2007ANL,Khler2016RobustMS,Pickup2007OvercomingRU,Zhang2012CommutabilityOB}, or marginalizing over them~\cite{Pickup2007OvercomingRU,Pickup2009BayesianMF,Tipping2002BayesianIS}. Alternatively, a number of approaches directly predict the HR image without simulating the image formation process. Chiang and Boult~\cite{Chiang2000EfficientSV} upsample and warp the input images to a common reference, before fusing them. Farsiu~\etal~\cite{Farsiu2003RobustSA} extend this approach with a robust regularization term. Takeda~\etal~\cite{Takeda2006RobustKR,Takeda2007KernelRF} proposed a kernel regression based approach for super-resolution. 
Wronski~\etal~\cite{Wronski2019HandheldMS} used the kernel regression technique to perform joint demosaicking and super-resolution. A few deep learning based solutions have also been proposed recently for MFSR, mainly focused on remote sensing applications~\cite{Deudon2020HighResnetRF,Kawulok2019DeepLF,Molini2020DeepSUMDN}. Bhat~\etal~\cite{Bhat2021DeepBS} propose a learned attention-based fusion approach for hand held burst super-resolution.
Haris~\etal~\cite{Haris2019RecurrentBN} propose a recurrent back-projection network for video super-resolution.

\parsection{Multi-Frame Denoising} In addition to the MFSR approaches discussed previously, a number of specialized multi-frame denoising approaches have also been proposed in the literature. Tico~\cite{Tico2008MultiframeID} performs block matching both within an image, as well as across the input images to perform denoising. \cite{Dabov2007VideoDB,Maggioni2011VideoDU,Maggioni2012VideoDD} extend the popular image denoising algorithm BM3D~\cite{Dabov2007ImageDB} to video. Buades~\etal~\cite{Buades2009ANO} estimate the noise level from the aligned images, and use a combination of pixel-wise mean and BM3D to denoise. 
Hasinoff~\etal~\cite{Hasinoff2016BurstPF} used a hybrid 2D/3D Wiener filter to denoise and merge burst images for HDR and low-light photography applications. 
Godard~\etal~\cite{Godard2018DeepBD} extend a single frame denoising network for multiple frames using a recurrent neural network. Mildenhall~\etal~\cite{Mildenhall2018BurstDW} employed a kernel prediction network (KPN) to obtain per-pixel kernels which are used to merge input images. The KPN approach was then extended by~\cite{Marinc2019MultiKernelPN} to predict multiple kernels, while~\cite{Xia2020BasisPN} introduced basis prediction networks to enable the use of larger kernels. 

\parsection{Deep Optimization-based image restoration} A number of deep learning based approaches~\cite{Kokkinos2018DeepID,Kokkinos2019IterativeRC,Zhang2020DeepUN,Zhang2017LearningDC} have posed image restoration tasks as an explicit optimization problem. 
The $P^3$~\cite{Venkatakrishnan2013PlugandPlayPF} and RED~\cite{Reehorst2019RegularizationBD} approaches provide a general framework for utilizing standard denoising methods as regularizers in optimization-based image restoration methods. 
Zhang~\etal~\cite{Zhang2017LearningDC} used the half quadratic splitting method to plug a deep neural-network based denoiser prior into model-based optimization methods. 
Kokkinos~\etal~\cite{Kokkinos2019IterativeRC} used a proximal gradient descent based framework to learn a regularizer network for burst photography applications. These prior works mainly focus on only learning the regularizer, while assuming that the data term (image formation process) is known and simple. Furthermore, the reconstruction error computation, as well as the error minimization are restricted to be in the image space. In contrast, our deep reparametrization approach allows jointly learning the imaging process as well as the priors, without restricting the image formation model to be simple or linear.   
\section{Method}

\begin{figure*}[t]
    \centering%
    \includegraphics[trim = 0 0 0 0, width=0.97\textwidth]{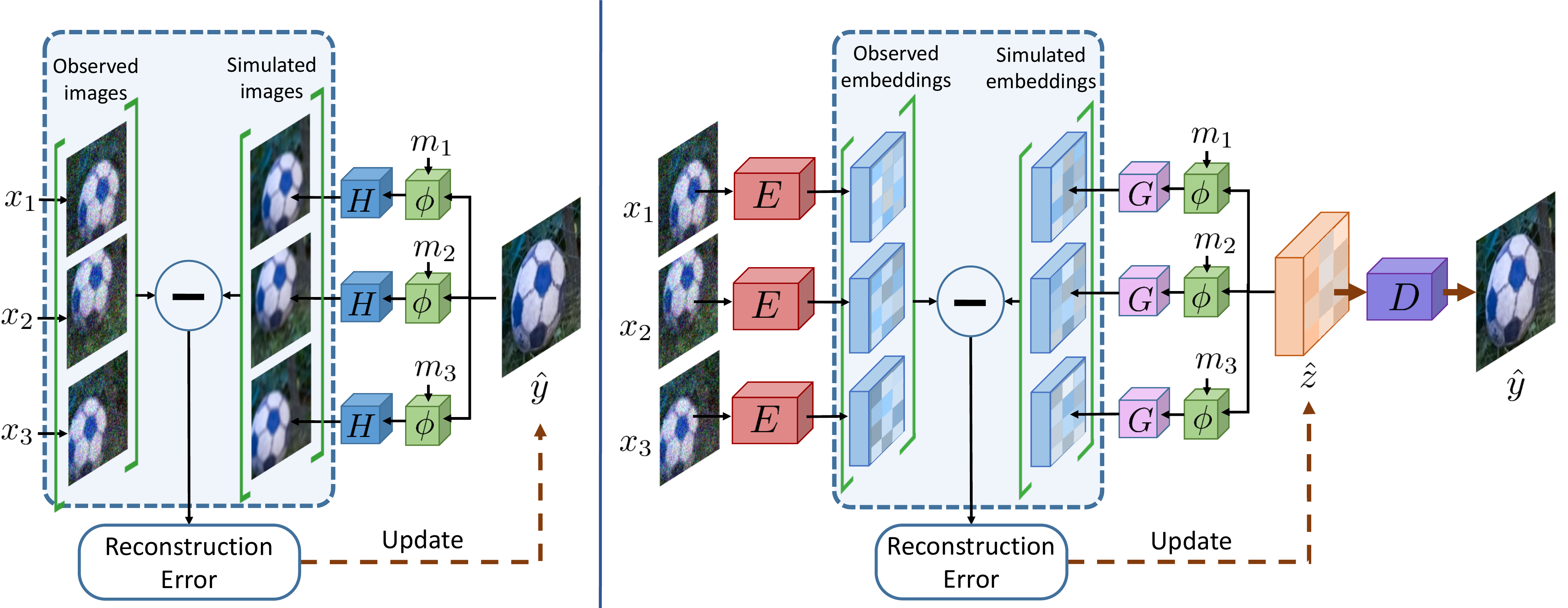}\vspace{0mm}
    \caption{\textbf{Left}: Classical multi-frame image restoration approaches minimize a reconstruction error~\eqref{eq:map} between the observed images $x_i$ and the simulated images $\h\left(\shift_{\motion_i}\left(\pred\right)\right)$ to obtain the output image $\pred$. \textbf{Right}: In contrast, we employ an encoder $\enc$ to compute the reconstruction error~\eqref{eq:ours_loss3} in a learned feature space. The reconstruction error is minimized \wrt a latent representation $\predenc$, which is then passed through the decoder $\dec$ to obtain the prediction $\pred$. }\vspace{-4mm}
    \label{fig:method_overview}
\end{figure*}

\subsection{Problem formulation}
\label{sec:problem_form}
In this work, we tackle the multi-frame (MF) image restoration problem. Given multiple images $\{x_i\}_{i=1}^{N}$, $x_i \in \reals^{h \times w \times c_\text{in}}$ of a scene, the goal is to merge information from these input images to generate a higher quality output $\pred \in \reals^{sh \times sw \times c_\text{out}}$. Here, $c_\text{in}$ and $c_\text{out}$ are the number of image channels, while $s$ is the super-resolution factor.
We consider a general scenario where the input images are either captured using a stationary or a hand held camera. The input and output images can either be in RAW or RGB format, depending on the end application.

One of the most successful paradigms to MF restoration and super-resolution in the literature~\cite{Bascle1996MotionDA,Farsiu2004MultiframeDA,Hardie1998HighResolutionIR,Peleg1987ImprovingIR} is to first model the image formation process,
\begin{equation}
\label{eq:image_formation}
    x_i = \h\big( \shift_{\motion_i}(\yc )\big) + \noise_i
\end{equation}
Here, $\yc$ is the underlying image and $\shift_{\motion_i}$ is the warping operation which accounts for scene motion $\motion_i$. The image degradation operator $\h$ models \eg camera blur, and the sampling process in camera. The observation noise $\noise_i \sim \noisedist$ is assumed to follow a given distribution $\noisedist$. The degradation operator $\h$ and the scene motion $\motion_i$ take different forms depending on the addressed task. For example, in the super-resolution task, $\h$ acts as the downsampling kernel.
Similarly, the scene motion $\motion_i$ can denote the parameters of an affine transformation, or represent a per-pixel optical flow in case of dynamic scenes. Note that the degradation operator $\h$ as well as the scene motion $\motion_i$ are unknown in general and need to be estimated.

Given the imaging model \eqref{eq:image_formation}, the original image $y$ is generally estimated by minimizing the error between each observed image $x_i$, and its simulated counterpart $\Bar{x}_i = \h\left( \shift_{\motion_i}\left(y\right)\right)$, using the maximum a posteriori (MAP) estimation technique. 
If the observation noise follows an i.i.d.\ Gaussian distribution, the MAP estimate $\pred$ is obtained as,
\begin{equation}
\label{eq:map}
    \pred = \argmin_{y} \sum_{i=1}^{N} \big\|x_i - \h\left(\shift_{\motion_i}(y)\right) \big\|_2^2 + \reg(y)
\end{equation}
Here, $\reg(y)$ is the regularization term that integrates prior knowledge about the original image $y$. 

The formulation \eqref{eq:map} provides a principled way of integrating information from multiple frames, leading to its popularity. However, it requires manually tuning the degradation operator $\h$ and the regularizer $\reg$, while also lacking the flexibility to generalize to more complex noise distributions. In this work, we propose a deep reparametrization of \eqref{eq:map} to address the aforementioned issues.

\subsection{Deep Reparametrization}
\label{sec:our_formulation}
We introduce a deep reparametrization that transforms the optimization problem \eqref{eq:map} into a learned deep feature space (see Figure~\ref{fig:method_overview}).
In this section, we will first derive our approach based on the reconstruction loss \eqref{eq:map}, and then discuss its advantages over the original image-space formulation. Our generalized deep image reconstruction objective is derived from \eqref{eq:map} in three steps, detailed next.

\parsection{Step 1}
We note that the first term in problem \eqref{eq:map} minimizes a $L_2$ distance $\|x_i - \Bar{x}_i\|_2$ between the observed image $x_i$ and the simulated image $\Bar{x}_i = \h\left( \shift_{\motion_i}\left(y\right)\right)$. Instead of limiting the objective to the squared error $\|x_i - \Bar{x}_i\|^2_2$ in image space, we learn a more general distance measure $d(x_i, \Bar{x}_i)$. We parametrize the metric $d$ by an encoder network $\enc$, to obtain image embeddings $\enc(x_i)  \in \reals^{\Tilde{h} \times \Tilde{w} \times c_\text{e}}$. The error $d(x_i, \Bar{x}_i)$ is then computed as the $L_2$ distance between the embeddings of the input image $x_i$ and the simulated image $\Bar{x}_i$, as $d(x_i, \Bar{x}_i) = \|\enc(x_i) - \enc(\Bar{x}_i)\|_2$. Thanks to the depth and non-linearity of the encoder $E$, the distance measure $d$ can represent highly flexible error metrics, more suitable for complex noise and error distributions.

\parsection{Step 2}
While the encoder $E$ maps the error computation to a deep feature space, the resulting objective is still minimized in the output image space $y$. As a second step, we therefore reparametrize the objective \eqref{eq:map} in terms of a latent deep representation $\predenc \in \reals^{\Tilde{s}\Tilde{h} \times \Tilde{s}\Tilde{w} \times c_\text{z}}$ of the image $y$. To this end, we introduce a decoder network $\dec$ that maps the latent representation $\predenc$ to the estimated image $y = \dec(\predenc)$. Since $\predenc$ is a direct parametrization of the target image $y$, we can optimize the objective \wrt $\predenc$ and predict the final image as $\pred = \dec(\hat{\predenc})$ once the optimal latent representation $\hat{\predenc}$ is found. The resulting objective is thus expressed as,
\begin{align}
\label{eq:ours_loss}
    L(\predenc) &= \sum_{i=1}^N \big\|\enc(x_i) - \enc  \circ \h  \circ \shift_{\motion_i} \circ \dec(z) \big\|_2^2  + \reg\left(\dec\left(\predenc\right)\right) \nonumber \\
    \pred &= \dec(\hat{\predenc}) \,,\quad \hat{\predenc} = \argmin_{z} L(z) \,.
\end{align}
Here, `$\circ$' denotes composition $f \circ g(\cdot) = f(g(\cdot))$ of two functions $f$, $g$.

Next, we assume the decoder $\dec$ to be equivariant \wrt the warping operation $\shift_{\motion_i}$. That is, the decoder and warping operation commute as $\shift_{\motion_i} \circ \dec = \dec \circ \shift_{\motion_i}$. In fact, if $\shift_{\motion_i}$ is solely composed of a translation, this condition is readily ensured by the translational equivariance of the CNN decoder $\dec$. For more complex motions, the equivariance condition still holds to a good approximation if the motion $\motion_i$ locally resembles a translation. This is generally the case for the considered burst photography settings, where the motion between frames are small to moderate. Furthermore, as for optical flow networks that also employ feature warping \cite{Ilg2017FlowNet2E,Sun2018PWCNetCF}, our decoder $\dec$ can learn to accommodate the desired warping equivariance through end-to-end training. By using the equivariance condition $\shift_{\motion_i} \circ \dec = \dec \circ \shift_{\motion_i}$ in \eqref{eq:ours_loss}, we obtain the objective,
\begin{equation}
\label{eq:ours_loss2}
    L(\predenc) \!=\! \sum_{i=1}^N \big\|\enc(x_i) - \underbrace{\enc  \circ \h \circ \dec}_G \circ \,\shift_{\motion_i}(z) \big\|_2^2  + \reg\left(\dec\left(\predenc\right)\right),
\end{equation}
which allows us to directly apply the warping $\shift_{\motion_i}$ on the latent representation $z$.

\parsection{Step 3}
As the final step, we focus on the degradation operator $\h$. In general, $\h$ is unknown and thus needs to be estimated or learned. Although it could be directly parametrized as a separate neural network, we propose a different strategy. By directly comparing \eqref{eq:map} and \eqref{eq:ours_loss2}, we interestingly find the role of $\h$ in \eqref{eq:map} replaced by the composition $G = \enc  \circ \h \circ \dec$ in \eqref{eq:ours_loss2}. Instead of learning the image space degradation map $\h$, we can thus directly parametrize its resulting deep feature space operator $G$. 
Here, $G$ can be seen as the feature space degradation operator, which is used to directly obtain the simulated image embedding $G\left(\shift_{\motion_i}(z)\right)$.
We thereby obtain the following objective, 
\begin{equation}
\label{eq:ours_loss3}
    L(\predenc) = \sum_{i=1}^N \left\|\enc(x_i) - G\big(\shift_{\motion_i}(z)\big) \right\|_2^2  + \regenc(\predenc)\,.
\end{equation}
In \eqref{eq:ours_loss3}, we have also introduced the latent space regularizer $\regenc = \reg \circ \dec$, which can similarly be parametrized directly in order to avoid invoking the decoder $\dec$ during the optimization process. 
Next, we will discuss the advantages of our deep reformulation \eqref{eq:ours_loss3} of \eqref{eq:map}, brought by each of the neural network modules $\enc$, $\dec$, and $G$.

\parsection{Encoder $\enc$} The encoder maps the input images $x_i$ to an embedding space $\enc(x_i)$, where the reconstruction error is defined. It can thus learn to transform complex noise distributions $\noisedist$ and other error sources, stemming from \eg inaccurate motion estimation $\motion_i$, to a feature space where it is better approximated as independent Gaussian noise. Our approach thus avoids strict assumptions imposed by the $L_2$ loss in image space through the flexibility of the encoder $\enc$.

\parsection{Decoder $\dec$} The minimization problem \eqref{eq:map} is often solved using iterative numerical methods, such as the conjugate gradient method. The convergence rate of such methods strongly depend on the conditioning of the objective. Since we optimize \eqref{eq:ours_loss2} \wrt a latent representation $z$ instead of the output image $y$, our decoder $\dec$ serves as a preconditioner, leading to faster convergence. Furthermore, while effective image space regularizers $\reg(y)$ are often complex \cite{Farsiu2004MultiframeDA,Gotoh2004DirectSA}, our latent parametrization $z$ allows for trivial regularizers $\regenc(\predenc)$. Similar to CNN-based single-image super-resolution approaches \cite{Dong2014LearningAD,Lai2017DeepLP,Lim2017EnhancedDR,Zhang2018ResidualDN}, the decoder $\dec$ also learns strong image priors which are applied during the prediction step $\pred = \dec(\hat{\predenc})$. Thanks to the regularizing effect of our decoder, we found it sufficient to simply set $\regenc(\predenc) = \lambda \|z\|^2_2$ where $\lambda$ is a learnable scalar.

\parsection{Feature degradation $G$} The image degradation operator $\h$ can be complex and non-linear in general, making it hard to solve the minimization problem \eqref{eq:map}. In our deep reformulation \eqref{eq:ours_loss3} of \eqref{eq:map}, the image degradation $\h$ is replaced by its feature space counterpart $G = \enc \circ \h \circ \dec$. Here, the encoder $E$ and decoder $D$ are deep neural networks, capable of learning highly non-linear mappings. These can
therefore learn a latent space where the degradation operation is approximately linear. That is, for a given image
degradation $H$, we can learn appropriate $G$, $E$ and $D$ such that $G \approx \enc \circ \h \circ \dec$ even in the case when $G$ is constrained to be linear. Consequently, we constrain $G$ to be a linear convolution filter, which is accommodated by the end-to-end learning of suitable $E$ and $D$ where such a linear relation holds. As a result, our optimization problem \eqref{eq:ours_loss3} is convex and can be easily optimized using efficient quadratic solvers.

We model the encoder $\enc$, decoder $\dec$, and degradation $G$ as convolutional neural networks. As detailed in Sec.~\ref{sec:training}, these networks are learned directly from data. But first, we propose a further generalization to our objective~\eqref{eq:ours_loss3}.

\subsection{Certainty Predictor}
\label{sec:certainty_pred}
In our formulation \eqref{eq:ours_loss3}, the reconstruction error for each frame, location, and feature channel are weighted equally. This is the correct model if the errors, often seen as observation noise, are identically distributed. In practice however, images are affected by heteroscedastic noise \cite{Healey1994RadiometricCC}, which varies spatially depending on the image intensity value. Furthermore, the reconstruction errors in \eqref{eq:map} and \eqref{eq:ours_loss3} are affected by the quality of the motion estimation $\motion_i$. In practical applications, the scene motion $\motion_i$ is unknown and needs to be estimated using \eg optical flow. As a result, the estimated $\motion_i$ may contain significant errors for certain regions, leading to sub-optimal results. In order to model these effects, we further introduce a certainty predictor module $\weightpred$.

Our certainty predictor aims to determine element-wise certainty values $v_i \in \reals^{\Tilde{h} \times \Tilde{w} \times c_\text{e}}$ for each element in the residual $\enc(x_i) - G\left(\shift_{\motion_i}\left(z\right)\right)$. Intuitively, image regions with higher noise or unreliable motion estimate $\motion_i$ should be given lower certainty weights, effectively reducing their impact in the MAP objective~\eqref{eq:ours_loss3}. The certainty values $v_i$ are computed using the image embeddings $\{\enc(x_j)\}_{j=1}^N$, motion estimate $\motion_i$, and the noise level $n_i$ (if available) as input. Our final optimization problem, including the certainty weights $v_i$ is then expressed as,
\begin{align}
\label{eq:ours_loss_w}
    L(z) =&  \sum_{i=1}^N \|v_i \cdot \left(\enc(x_i) - G\left(\shift_{\motion_i}\left(z\right)\right)\right) \|_2^2  + \lambda\|\predenc\|^2_2 \nonumber\\
    &\text{where}\quad v_i = \weightpred\big(\{\enc(x_j)\}_{j=1}^N, \motion_i, n_i\big) \,.
\end{align}
In relation to the MAP estimation \eqref{eq:map}, the certainty weights correspond to an estimate of the inverse standard deviation $v_i = \frac{1}{\sigma_i}$ of the encoded observations $\enc(x_i)$.

\subsection{Optimization}
\label{sec:optimization}
To ensure practical inference and training, it is crucial that our objective \eqref{eq:ours_loss_w} can be minimized efficiently.
Furthermore, in order to learn our network components end-to-end, the optimization solver itself needs to be differentiable. Due to the linearity of warp operator $\shift_{\motion_i}$ and the choice of linear feature degradation $G$, our objective $L(z)$ is a linear least-squares problem, which can be addressed with standardized techniques. In particular, we employ the steepest-descent algorithm, which can be seen as a simplification of the Conjugate Gradient~\cite{CGpain}. 
Both algorithms have been previously employed in classical MFIR approaches~\cite{Bascle1996MotionDA,Hardie1998HighResolutionIR}, and more recently in deep optimization-based few-shot learning approaches~\cite{bhat2019learning,bhat2020learning,Tripathi2020FewShotCB}.


The steepest-descent algorithm performs an optimal line search $\alpha^j = \argmin_\alpha L(z^j - \alpha g^j)$ in the gradient $g^j = \nabla L(z^j)$ direction to update the iterate $z^{j+1} = z^j - \alpha^j g^j$. Since the problem is quadratic, simple closed-form expressions can be derived for both the gradient $g^j$ and step length $\alpha^j$. For our model \eqref{eq:ours_loss_w}, the complete algorithm is given by, 
\begin{align}
    \label{eq:sd}
    &g^j = -2\sum_{i=1}^N \shift_{\motion_i}^{\text{T}} G \conv^{\text{T}} \left(v_i^2 \!\cdot\! \left(\enc(x_i) - G \conv \shift_{\motion_i}\!(z^j)\right)\right) + 2\lambda z^j \nonumber\\
    &\alpha^j = \frac{\|g^j\|_2^2}{\sum_{i=1}^N  2\|v_i \cdot \left(G \conv \shift_{\motion_i}\left(g^i\right)\right) \|_2^2 + 2\lambda \|g^j\|_2^2} \\
    &z^{j+1} = z^j - \alpha^j g^j \,. \nonumber
\end{align}
Here, $\conv$, $\conv^{\text{T}}$, and $\cdot$ denote the convolution, transposed convolution, and element-wise product, respectively. Further, $\shift_{\motion_i}^{\text{T}}$ is the transposed warp operator. A detailed derivation is provided in the supplementary material. Note that both the gradient $g^j$ and step length $\alpha^j$ can be implemented using standard differentiable neural network operations. 

To further improve convergence speed, we learn an initializer $\modelinit$ which predicts the initial latent encoding $z^0 = \modelinit(\enc(x_1))$ using the embedding of the first image $x_1$. Our approach then proceeds by iteratively applying $K_\text{SD}$ steepest-descent iterations \eqref{eq:sd}. Due to the fast convergence provided by the steepest-descent steps, we found it sufficient to only use $K_\text{SD} = 3$ iterations. 
By unrolling the iterations, our optimization module can be represented as a feed-forward network $\solver_{G, \weightpred, \modelinit}$ that predicts the optimal encoding $\hat{z}$. Our complete inference procedure is then expressed as,
\begin{equation}
    \label{eq:inference}
    \pred = \dec\left(\solver_{G, \weightpred, \modelinit}\left(\big\{\big(\enc(x_i), m_i\big)\big\}_{i=1}^N\right) \right)  
\end{equation}
In the next section, we will describe how all the components in our architecture can be directly learned end-to-end.

\subsection{Training}
\label{sec:training}

Our entire MFIR network is trained end-to-end from data in a straightforward manner, without enforcing any additional constraints on the individual components. We use a training dataset $\mathcal{D} = \{(\{x_i^k\}_{i=1}^N, y^k)\}$ consisting of input-target pairs. For each input $\{x_i^k\}_{i=1}^N$, we obtain the prediction $\pred^k$ using \eqref{eq:inference}. The network parameters for each of our components $\enc$, $G$, $\weightpred$, $\modelinit$, and $\dec$ are then learned by minimizing a prediction error $\ell(y^k, \pred^k)$ over the training dataset $\mathcal{D}$ using \eg stochastic gradient descent. In this work, we use the popular $L_1$ loss $\ell(y, \pred) = \|y - \pred\|_1$.

\section{Applications}
We describe the application of our approach to RAW burst super-resolution, and burst denoising tasks. A detailed description is provided in the supplementary material.

\subsection{RAW Burst Super-resolution}
\label{sec:app_sr}
Here, the method is given a set of RAW bayer images captured successively from a hand held camera. 
 The task is to exploit these multiple shifted observations to generate a denoised, demosaicked, higher-resolution output. In this setting, the image degradation $\h$ can be seen as a composition of camera blur, decimation, sampling, and mosaicking operations. Next, we briefly describe our architecture.

\parsection{Encoder $\enc$} The encoder packs each 2 $\times$ 2 block in the input RAW image along the channel dimension to obtain a 4 channel input.  This is then passed through an initial conv.\ layer followed by a series of residual blocks \cite{He2016DeepRL} with ReLU activations and without BatchNorm~\cite{Ioffe2015BatchNA}. A final conv.\ layer predicts a $256$-dimensional encoding of the input image.

\parsection{Operator $G$} We use a conv.\ layer with stride $\Tilde{s}$ as our feature-space degradation $G$. The stride $\Tilde{s}$ corresponds to the downsampling factor of $G$. Note that this downsampling need not be the same as the downsampling factor $s$ of the image degradation $\h$. Our latent representation $\predenc$ can encode higher-resolution information in the channel dimension, enabling use of a smaller $\Tilde{s}$ for efficiency.
We empirically observed that it is sufficient to set $\Tilde{s} = 2$ and perform the remaining upsampling by factor $s/\Tilde{s}$ in our decoder $\dec$. 

\parsection{Decoder $\dec$} Our decoder consists of a series of residual blocks (same type as in $E$), followed by upsampling by a factor of $s / \Tilde{s}$ using sub-pixel convolution~\cite{Shi2016RealTimeSI}. The upsampled feature map is passed through additional residual blocks, followed by a final conv.\ layer to obtain $\pred$.

\parsection{Motion Estimation} We compute the motion $\motion_i$ between each input image $x_i$ and a reference image $x_1$ as pixel-wise optical flow in order to be robust to small object motions in the scene. Specifically, we use a PWCNet~\cite{Sun2018PWCNetCF} trained by the authors on the synthetic FlyingChairs~\cite{Dosovitskiy2015FlowNetLO}, FlyingThings3D~\cite{Mayer2016ALD}, and MPI Sintel~\cite{Butler2012ANO} datasets.

\parsection{Certainty predictor $\weightpred$} We uses three sources of information in order to predict the certainty $v_i$: i) The encoding $\enc(x_i)$ which provides information about local image structure \eg presence of an edge, texture \etc. ii) The residual $\enc(x_i) - \shift_{\motion_i}(\enc(x_1))$ between the encoding of $i$-th image $x_i$ and the reference image encoding $\enc(x_1)$ warped to $i$-th image, which can indicate possible alignment failures, and iii) The sub-pixel sampling location $\motion_i$ mod $1$ of the pixels in the $i$-th image. These three entities are passed through a residual network to obtain the certainty $v_i$ for image $x_i$.

\subsection{Burst Denoising}
\label{sec:app_denoise}
Given a burst of noisy images, the aim of burst denoising is to generate a clean output image. In general, burst denoising requires filtering over both the temporal and spatial dimensions. While the classical MAP formulation \eqref{eq:map} accommodates the latter by specially designed regularizers, our approach can learn spatial filtering through two mechanisms. First, the encoder $E$ and decoder $D$ networks allow effective spatial aggregation. Second, our certainty predictor can predict both frame-wise and spatial (through channel-dimension encodings) aggregation weights. Following~\cite{Marinc2019MultiKernelPN,Mildenhall2018BurstDW,Xia2020BasisPN}, we consider a burst denoising scenario where an estimate of per-pixel noise variance $n_i$ is available. In practice, such an estimate is available from the exposure parameters reported by the camera. Next, we briefly detail our network architecture employed for this task.

\parsection{Encoder $\enc$} We concatenate the image $x_i$ and the noise estimate $n_i$ and pass it through a residual network to obtain the noise conditioned image encodings $\enc(x_i, n_i)$

\parsection{Operator $G$} We use a conv.\ layer as our operator $G$.

\parsection{Decoder $\dec$} Our decoder consists of a series of residual blocks, followed by a final conv.\ layer which outputs $\pred$.

\parsection{Motion Estimation} We use a similar strategy as employed in Sec.~\ref{sec:app_sr} to estimate the motion between images.

\parsection{Certainty predictor $\weightpred$} We use a similar certainty predictor as employed in Sec.~\ref{sec:app_sr}, with a minor modification. We input the noise estimate $n_i$ directly to $\weightpred$ as to condition our minimization problem \eqref{eq:ours_loss_w} on the input noise level.

\section{Experiments}
We perform comprehensive evaluation of our approach on RAW burst super-resolution and burst denoising tasks. Detailed results are provided in the supplementary material.

\begin{figure*}[t]
    \centering%
    \includegraphics[trim = 0 0 0 0, width=0.99\textwidth]{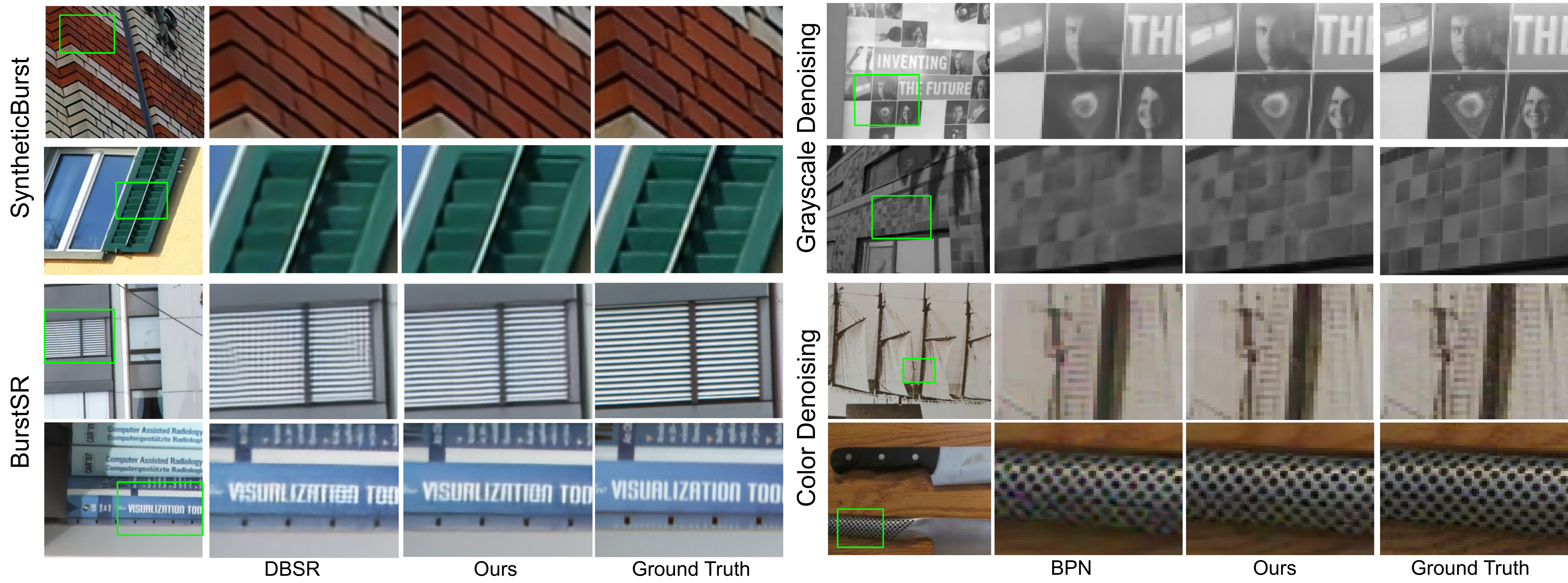}\vspace{0mm}
    \caption{Qualitative comparison of our approach with the previous state-of-the-art methods DBSR~\cite{Bhat2021DeepBS} and BPN~\cite{Xia2020BasisPN} on RAW burst super-resolution (first four columns) and burst denoising (last four columns) tasks.}\vspace{-4mm}
    \label{fig:qual_joint}
\end{figure*}

\subsection{RAW Burst Super-Resolution}
\label{sec:exp_sr}
Here, we evaluate our approach on the RAW burst super-resolution task. Our experiments are performed on the SyntheticBurst dataset, and the BurstSR dataset, both introduced in~\cite{Bhat2021DeepBS}. The SyntheticBurst dataset consists of synthetically generated RAW bursts, each containing 14 images. The bursts are generated by applying random translations and rotations to a sRGB image, and converting the shifted images to RAW format using an inverse camera pipeline~\cite{Brooks2019UnprocessingIF}. The BurstSR dataset, on the other hand, contains real-world bursts captured using a hand held smartphone camera, along with a high-resolution ground truth captured using a DSLR camera. Since the input bursts and HR ground truth are captured using different cameras, there are spatial and color mis-alignments between the two, posing additional challenges for both training and evaluation. We perform super-resolution by a factor $s = 4$ in all our experiments.

\parsection{Training details} For evaluation on the SyntheticBurst dataset, we train our model on synthetic bursts generated using sRGB images from the Zurich RAW to RGB~\cite{ignatov2020replacing} training set. We use a fixed burst size $N = 14$ during our training. Our model is trained for $500$k iterations, with a batch size of $16$, using the ADAM~\cite{Kingma2015AdamAM} optimizer. The model trained on the synthetic data is then additionally fine-tuned for $40$k iterations on the BurstSR training set for evaluation on the BurstSR val set. In order to handle the mis-alignments between the input and the ground truth in BurstSR dataset, we perform a spatial and color alignment of the network prediction to the ground truth, using the strategy employed in~\cite{Bhat2021DeepBS}, before computing the prediction error. 

\begin{table}[!t]
	\centering\vspace{-1mm}
	\resizebox{\columnwidth}{!}{%
		\begin{tabular}{l|ccc|ccc|c}
\toprule
&\multicolumn{3}{c|}{SyntheticBurst} & \multicolumn{3}{c|}{BurstSR} & \\
&PSNR$\uparrow$&LPIPS $\downarrow$&SSIM$\uparrow$&PSNR$\uparrow$&LPIPS $\downarrow$&SSIM$\uparrow$&Time (s)\\\midrule
SingleImage & 36.86 & 0.113& 0.919& 46.60 & 0.039 & 0.979&\first{0.02}\\
HighResNet~\cite{Deudon2020HighResnetRF}& 37.45  & 0.106  & 0.924  & 46.64 & 0.038 & 0.980&\second{0.11}\\
DBSR~\cite{Bhat2021DeepBS} & \second{40.76} & \second{0.053} & \second{0.959}& \second{48.05} & \second{0.025} & \second{0.984}&0.24\\
\textbf{Ours}&  \first{41.56}  & \first{0.045} & \first{0.964}& \first{48.33}& \first{0.023} & \first{0.985}&0.40\\
\bottomrule
\end{tabular}

	}\vspace{1mm}%
	\caption{Comparison on the SyntheticBurst and real-world BurstSR validation dataset from~\cite{Bhat2021DeepBS}.}
	\label{tab:synburst}%
	\vspace{-6mm}
\end{table}

\parsection{Results} We compare our approach with the recently introduced DBSR~\cite{Bhat2021DeepBS} that employs a deep network with an attention-based fusion of input images. Our approach employs the same optical flow estimation network as DBSR. We also compare with HighResNet~\cite{Deudon2020HighResnetRF}, and a CNN-based single-image baseline consisting of only our encoder and decoder modules. All models are trained using the same training settings as our approach, and evaluated using all available burst images ($N = 14$). The results on the SyntheticBurst dataset containing $300$ bursts, in terms of PSNR, SSIM~\cite{Wang2004ImageQA}, and LPIPS~\cite{Zhang2018TheUE} are shown in Tab.~\ref{tab:synburst}. All metrics are computed in linear image space. Our approach, minimizing a feature-space reconstruction error, obtains the best results, outperforming DBSR by $+0.80$ dB in PSNR. We also report results on the real-world BurstSR val set containing $882$ bursts, using the evaluation strategy described in~\cite{Bhat2021DeepBS} to handle the spatial and color mis-alignments. 
Our approach obtains promising results, outperforming DBSR by $+0.28$ dB in PSNR. These results demonstrate that our deep reparametrization of the classical MAP formulation generalizes to real-world degradation and noise. The computation time required to process a burst containing 14 RAW images to generate a 1896 × 1080 RGB output is also reported in Tab.~\ref{tab:synburst}. A qualitative comparison is provided in Fig.~\ref{fig:qual_joint}.

\begin{table}[!t]
	\centering\vspace{-1mm}
	\resizebox{0.99\columnwidth}{!}{%
		\begin{tabular}{lccccc}
\toprule
&Gain $\propto$ 1&Gain $\propto$ 2&Gain $\propto$ 4&Gain $\propto$ 8& Average\\\midrule
HDR+~\cite{Hasinoff2016BurstPF}& 31.96  &  28.25 & 24.25  & 20.05&26.13\\
BM3D~\cite{Dabov2007ImageDB}& 33.89  &  31.17 & 28.53  & 25.92&29.88\\
NLM~\cite{Buades2005ANA}& 33.23  &  30.46 & 27.43  & 23.86&28.75\\
VBM4D~\cite{Maggioni2012VideoDD}& 34.60  &  31.89 & 29.20  & 26.52&30.55\\
SingleImage& 35.16  &  32.27 & 29.34  & 25.81&30.65\\
KPN~\cite{Mildenhall2018BurstDW}&36.47&33.93&31.19&27.97&32.39\\
MKPN~\cite{Marinc2019MultiKernelPN}&36.88&34.22&31.45&28.52&32.77\\
BPN~\cite{Xia2020BasisPN}&38.18&35.42&32.54&\second{29.45}&33.90\\
\textbf{Ours}&\first{39.37}&\first{36.51}&\first{33.38}&\first{29.69}&\first{34.74}\\
\textbf{Ours$^\dag$}& \second{39.10} & \second{36.14} & \second{32.89} & 28.98&\second{34.28}\\
\bottomrule
\end{tabular}
	}\vspace{1mm}%
	\caption{Comparison of our method with prior approaches on the grayscale burst denoising set~\cite{Mildenhall2018BurstDW} in terms of PSNR. Results for the first four methods are from~\cite{Mildenhall2018BurstDW}, while the results for MKPN are from~\cite{Xia2020BasisPN}. Our approach obtains the best results, outperforming the previous state-of-the-art method BPN on all noise levels.
	}
	\label{tab:kpn_grayscale}%
	\vspace{-4mm}
\end{table}

\subsection{Burst Denoising}
\label{sec:exp_denoise}
We evaluate our approach on the grayscale and color burst denoising datasets introduced in~\cite{Mildenhall2018BurstDW} and~\cite{Xia2020BasisPN}, respectively. Both datasets are generated synthetically by applying random translations to a base image. The shifted images are then corrupted by adding heteroscedastic Gaussian noise~\cite{Healey1994RadiometricCC} with variance $\sigma_r^2 + \sigma_s x$. Here $x$ is the clean pixel value, while $\sigma_r$ and $\sigma_s$ denote the read and shot noise parameters, respectively. During training, the noise parameters $(\log(\sigma_r), \log(\sigma_s))$ are sampled uniformly in the log-domain from the range $\log(\sigma_r) \in [-3, -1.5]$ and $\log(\sigma_s) \in [-4, -2]$. The networks are then evaluated on 4 different noise gains $(1, 2, 4, 8)$, corresponding to noise parameters $(-2.2, -2.6)$, $(-1.8, -2.2)$, $(-1.4, -1.8)$, and $(-1.1, -1.5)$, respectively. Note that the noise parameters for the highest noise gain (Gain $\propto$ 8) are unseen during training. Thus, performance on this noise level can indicate the generalization of the network to unseen noise. The noise parameters $(\log(\sigma_r), \log(\sigma_s))$ are assumed to be known both during training and testing, and can be utilized to estimate per-pixel noise variance.

\parsection{Training details} Following~\cite{Mildenhall2018BurstDW}, we use the images from the Open Images~\cite{openimages} training set to generate synthetic bursts. We train on bursts containing $N = 8$ images with resolution $128 \times 128$. Our networks are trained using the ADAM~\cite{Kingma2015AdamAM} optimizer for $150$k and $300$k iterations for the grayscale and color denoising tasks, respectively. The entire training takes less than $40$h on a single Nvidia V100 GPU. 

\begin{table}[!t]
	\centering\vspace{-1mm}
	\resizebox{0.99\columnwidth}{!}{%
		\begin{tabular}{lccccc|c}
\toprule
&Gain $\propto$ 1&Gain $\propto$ 2&Gain $\propto$ 4&Gain $\propto$ 8&Average&Runtime (s)\\\midrule
SingleImage& 37.94  &  34.98 & 31.74  & 28.03 & 33.17 & \first{0.005}\\
KPN~\cite{Mildenhall2018BurstDW}&38.86&35.97&32.79&30.01 & 34.41 & -\\
BPN~\cite{Xia2020BasisPN}&40.16&37.08&33.81&31.19 & 35.56 & 0.328\\
\textbf{Ours}&\first{42.21}&\first{39.13}&\first{35.75}&\first{32.52}& \first{37.40} & 0.198\\
\textbf{Ours$^\dag$}&\second{41.90}&\second{38.85}&\second{35.48}&\second{32.29}& \second{37.13} & \second{0.046}\\
\bottomrule
\end{tabular}
	}\vspace{1mm}%
	\caption{Comparison with previous methods on the color burst denoising set~\cite{Xia2020BasisPN} in terms of PSNR. The results for KPN are from~\cite{Xia2020BasisPN}. Our approach outperforms BPN on all four noise levels.}
	\label{tab:kpn_color}%
	\vspace{-4mm}
\end{table}

\parsection{Results} We compare our approach with the recent kernel prediction based approaches KPN~\cite{Mildenhall2018BurstDW}, MKPN~\cite{Marinc2019MultiKernelPN}, and BPN~\cite{Xia2020BasisPN}. Since our motion estimation network (PWCNet) is trained on external synthetic data, we include a variant of our approach, denoted as Ours$^\dag$, using a custom optical flow network. Our flow network is jointly trained with the rest of the architecture using a photometric loss, without any extra supervision or data.   
We also include results for the popular denoising algorithms~\cite{Buades2005ANA,Dabov2007ImageDB,Maggioni2012VideoDD} based on non-local filtering, the multi-frame HDR+ method~\cite{Hasinoff2016BurstPF}, as well as a single image baseline consisting of only our encoder and decoder.
The results over the $73$ bursts from the grayscale burst denoising dataset~\cite{Mildenhall2018BurstDW}, are shown in Tab.~\ref{tab:kpn_grayscale}. Our approach sets a new state-of-the-art, outperforming the previous best method BPN~\cite{Xia2020BasisPN} on all four noise levels. Ours$^{\dag}$ employing a custom flow network also obtains promising results, outperforming BPN on three out of four noise levels.

We also evaluate our approach on the recently introduced color burst denoising dataset~\cite{Xia2020BasisPN} containing $100$ bursts. The results, along with the computation time for processing a $1024 \times 768$ resolution burst, are shown in Tab.~\ref{tab:kpn_color}. Further qualitative comparison is provided in Fig.~\ref{fig:qual_joint}. As in the grayscale set, our approach obtains the best results, significantly outperforming the previous best method BPN. Ours$^{\dag}$ employing a custom flow network also outperforms BPN by over $1.5$ dB in average PSNR, while operating at a significantly higher speed. Furthermore, note that unlike BPN and KPN that are restricted to operate on fixed-size bursts, our approach can operate with bursts of any size, providing additional flexibility for practical applications. 

\subsection{Ablation Study}
\label{sec:exp_abl}
Here, we analyse the impact of key components in our formulation.
The experiments are performed on the SyntheticBurst super-resolution dataset~\cite{Bhat2021DeepBS} and the grayscale burst denoising dataset~\cite{Mildenhall2018BurstDW}. We train different variants of our approach, with and without the encoder $\enc$, decoder $\dec$, and the certainty predictor $\weightpred$. This is achieved by replacing the encoder/decoder by an identity function, and setting certainty weights $v_i$ to all ones, when applicable. In order to ensure fairness, we employ a deeper decoder when not utilizing an encoder, and vice versa. For training on the SyntheticBurst dataset, we employ a shorter training schedule with $100$k iterations. The mean PSNR on the SyntheticBurst set, as well as the mean PSNR over all four noise levels in the grayscale denoising set are provided in Tab.~\ref{tab:joint_ablation}.

Minimizing the reconstruction error directly in the input image space (MAP estimate~\eqref{eq:map}) leads to poor results on both super-resolution and denoising tasks (a). Note that unlike in the classical MAP based approaches, the degradation operator $\h$ is still learned in this case. 
The performance of the image space formulation is improved by employing our certainty predictor (b). The improvement is more prominent in the burst denoising task, where the certainty values allow handling varying noise levels. 
Our variants employing only the encoder (c), or the decoder (d) module obtain better performance, thanks to the increased modelling capability provided by the use of deep networks. 
The certainty predictor $\weightpred$ provides additional improvements, even when employed together with the encoder (f) or the decoder (g).
Removing either of the three components $\enc$, $\dec$, or $\weightpred$ (g)-(e) from our final version leads to a decrease in performance, demonstrating that each of these components are crucial. The performance decrease is much larger in the super-resolution task due to the more complex image degradation process.

\begin{table}[!t]
	\centering\vspace{-1mm}
	\resizebox{0.9\columnwidth}{!}{%
		\begin{tabular}{l|ccc|cc|cc}
\toprule
& &  &   & \multicolumn{2}{c|}{SyntheticBurst} &  \multicolumn{2}{c}{Denoising} \\
& $\enc$ & $\dec$  & $\weightpred$  & PSNR & $\Delta$PSNR & PSNR & $\Delta$PSNR\\\midrule
(a) & & & & 31.91 & -7.91 & 28.06 & -6.68\\
(b) & & & \checkmark & 33.85 & -5.97 & 33.00 & -1.74\\
(c) & \checkmark & & & 36.71 & -3.11 & 33.46 & -1.28\\
(d) & & \checkmark & & 38.12 & -1.70 & 32.99 & -1.75\\
(e) & \checkmark & \checkmark & & 38.36 & -1.46 & 34.54 & -0.20\\
(f) & \checkmark &  & \checkmark &  38.44 & -1.38 & 33.69 & -1.05\\
(g) & & \checkmark & \checkmark & 38.63 & -1.19 & 34.56 & -0.18\\
(h) & \checkmark & \checkmark & \checkmark & 39.82  &  & 34.74 & \\

\bottomrule
\end{tabular}
	}\vspace{1mm}%
	\caption{Impact of our encoder $\enc$, decoder $\dec$, and certainty predictor $\weightpred$ modules on SyntheticBurst~\cite{Bhat2021DeepBS} and grayscale denoising~\cite{Mildenhall2018BurstDW} datasets. $\Delta$PSNR denotes difference with our final model (h).}
	\label{tab:joint_ablation}%
	\vspace{-4mm}
\end{table}

\section{Conclusion}
We propose a deep reparametrization of the classical MAP formulation for multi-frame image restoration. Our approach minimizes the MAP objective in a learned deep feature-space, \wrt a latent representation of the output image. Crucially, our deep reparametrization allows learning complex image formation processes directly in latent space, while also integrating learned image priors into the prediction. We further introduce a certainty predictor module to provide robustness to \eg alignment errors. Our approach obtains state-of-the-art results on RAW burst super-resolution as well as burst denoising tasks.

\noindent\textbf{Acknowledgments}: This work was supported by a Huawei Technologies Oy (Finland) project, the ETH Z\"urich Fund (OK), an Amazon AWS grant, and Nvidia.

{\small
\bibliographystyle{ieee_fullname}
\bibliography{references}
}

\clearpage
\onecolumn

\setcounter{section}{0}
\renewcommand{\thesection}{\Alph{section}}

\begin{center}
	\textbf{\large Supplementary Material}
\end{center}

We provide additional details and analysis of our approach in this supplementary material. Section~\ref{sec:derivation} provides a derivation of the closed-form expressions for the steepest-descent steps \eqref{eq:sd}. The linearity of the warping operator is discussed in Section~\ref{sec:warp}. Our entire inference pipeline is detailed in Section~\ref{sec:inference_pipeline}. The network architectures employed for the RAW burst super-resolution and burst denoising tasks are described in detail in Section~\ref{sec:arch_details}. An analysis of our certainty predictor $W$ is provided in Section~\ref{sec:certainty_pred_supp}. Section~\ref{sec:ablation_supp} contains an additional ablative analysis of our approach. Further qualitative comparison on the burst super-resolution and denoising datasets are provided in Section~\ref{sec:qual_supp}.

\section{Derivation of Steepest-Descent Steps}
\label{sec:derivation}
In this section, we derive the closed-form expressions for the gradient $g^j = \nabla L (z^j)$ of loss \eqref{eq:ours_loss_w}, as well as the steepest-descent step lengths $\alpha^j$. Our optimization objective \eqref{eq:ours_loss_w} is here restated as,
\begin{align}
\label{eq:ours_loss_w_re}
    L(z) =&  \sum_{i=1}^N \|r_i \|_2^2  + \lambda\|\predenc\|^2_2 \\
    &\text{where}\quad r_i = v_i \cdot \left(\enc(x_i) - G \conv \shift_{\motion_i}\left(z\right)\right) \,. \nonumber
\end{align}
In the following derivation, we will interchangeably treat the entities in~\eqref{eq:ours_loss_w_re} either as 3D feature maps or as corresponding vectors. By using the chain rule, the gradient of~\eqref{eq:ours_loss_w_re} is computed as,
\begin{align}
\label{eq:grad_l}
    g = \nabla L(z) &= \sum_{i=1}^N 2\bigg[\frac{\partial{r_i}}{\partial{z}}\bigg]^\text{T}r_i  + 2\lambda \predenc \\
    &= \sum_{i=1}^N -2\bigg[\frac{\partial{G \conv \shift_{\motion_i}\left(z\right)}}{\partial{z}}\bigg]^\text{T} (v_i \cdot r_i)  + 2\lambda \predenc \\
    &= \sum_{i=1}^N -2\bigg[\frac{\partial{G \conv \shift_{\motion_i}\left(z\right)}}{\partial{\shift_{\motion_i}\left(z\right)}} \frac{\partial{ \shift_{\motion_i}\left(z\right)}}{\partial{z}}\bigg]^\text{T} (v_i \cdot r_i)  + 2\lambda \predenc \\
    &= \sum_{i=1}^N -2\bigg[\frac{\partial{\shift_{\motion_i}\left(z\right)}}{\partial{z}}\bigg]^\text{T} \bigg[\frac{\partial{G \conv \shift_{\motion_i}\left(z\right)}}{\partial{\shift_{\motion_i}\left(z\right)}}\bigg]^\text{T}(v_i \cdot r_i)  + 2\lambda \predenc \\
    &= \sum_{i=1}^N -2 \shift_{\motion_i}^\text{T} G \conv^\text{T}(v_i \cdot r_i)  + 2\lambda \predenc \\
    \label{eq:grad_l_final}
    &= -2 \sum_{i=1}^N \shift_{\motion_i}^\text{T} G \conv^\text{T}\left(v_i^2 \cdot \left(\enc(x_i) - G \conv \shift_{\motion_i}\left(z\right)\right) \right)  + 2\lambda \predenc
\end{align}
Here, $G \conv^\text{T}$ denotes the transpose of the convolution operator $u \mapsto G \conv u$, which is the same as the transpose of the Jacobian $\frac{\partial{G \conv u}}{\partial{u}}$. Similarly, $\shift_{\motion_i}^\text{T}$ denotes the transpose of the linear warp operator $z \mapsto \shift_{\motion_i}(z)$, which corresponds to the transpose of the Jacobian $\frac{\partial{\shift_{\motion_i}(z)}}{\partial{z}}$.

Our step length $\alpha^j$ is computed by performing an optimal line search $\alpha^j = \argmin_\alpha L(z^j - \alpha g^j)$ in the gradient direction $g^j$. Since our loss \eqref{eq:ours_loss_w_re} is convex, it has a unique global minima, which is be obtained by solving for the stationary point $\frac{\text{d}L(z^j - \alpha g^j)}{\text{d}\alpha} = 0$. By setting $u = z^j - \alpha g^j$ and applying chain rule, we get
\begin{align}
\label{eq:step_l}
0 = \frac{\text{d}L(u)}{\text{d}\alpha} &= \bigg[\frac{\text{d}u}{\text{d}\alpha}\bigg]^\text{T} \nabla L(u) \\
&= [-g^j]^\text{T} \bigg(-2 \sum_{i=1}^N \shift_{\motion_i}^\text{T} G \conv^\text{T}\left(v_i^2 \cdot \left(\enc(x_i) - G \conv \shift_{\motion_i}\left(u\right)\right) \right)   + 2\lambda u \bigg) \\
  &= 2[g^j]^\text{T} \sum_{i=1}^N \shift_{\motion_i}^\text{T} G \conv^\text{T}\left(v_i^2 \cdot \left(\enc(x_i) - G \conv \shift_{\motion_i}\left(z^j - \alpha g^j\right)\right) \right)  + 2\lambda [-g^j]^\text{T} (z^j - \alpha g^j) \\
  \label{eq:step_l_11}
  &= [-g^j]^\text{T} g^j  + 2 \sum_{i=1}^N \alpha[g^j]^\text{T} \shift_{\motion_i}^\text{T} G \conv^\text{T}\left(v_i^2 \cdot G \conv \shift_{\motion_i}\left( g^j\right) \right) + 2\lambda \alpha [g^j]^\text{T} g^j \\
  \label{eq:step_l_12}
  &= -\|g^j\|^2 + 2 \alpha   \sum_{i=1}^N \|v_i \cdot G \conv \shift_{\motion_i}\left( g^j\right)\|^2 + 2 \lambda \alpha \|g^j\|^2
\end{align}
In equation \eqref{eq:step_l_11}, we have utilized the closed-form expression \eqref{eq:grad_l_final} for $g^j$, while also exploiting the linearity of the warp operator $\shift$. Using \eqref{eq:step_l_12}, the step length $\alpha$ is obtained as,
\begin{equation}
    \alpha^j = \frac{\|g^j\|^2}{\sum_{i=1}^N 2\|v_i \cdot G \conv \shift_{\motion_i}\left( g^j\right)\|^2 + 2 \lambda \|g^j\|^2}
\end{equation}
\section{Linearity of the Warping Operator}
\label{sec:warp}
In our work, we assume that the warping operation $\phi_m(x)$ is linear. Note that this assumption holds even in the most general case, \ie where the scene motion is given by a pixel-wise optical flow, represented as a pixel-to-pixel mapping $m: \reals^2 \rightarrow \reals^2$. 
Let $x: \reals^2 \rightarrow \reals^d$ be a continuous representation of a $d$-dimensional feature map (obtained by \eg bilinear interpolation, which is itself linear). Then the warping operator can be conveniently expressed as a function composition $\phi_m(x) = x \circ m$, \ie $\phi_m(x)(p) = x(m(p))$ for any pixel location $p \in \reals^2$. The warping operator $\phi_m$ is hence linear for any motion $m$ since $\phi_m(a x_1 + b x_2) = (a x_1 + b x_2) \circ m = a x_1 \circ m + b x_2 \circ m = a \phi_m(x_1) + b \phi_m(x_2)$ for any scalars $a,b \in \reals$ and feature maps $x_1, x_2$.
\section{Inference Pipeline}
\label{sec:inference_pipeline}
Here, we detail the inference pipeline used by our multi-frame image restoration approach. Our approach minimizes the feature space reconstruction loss \eqref{eq:ours_loss_w} in the main paper to fuse information from the input images. The entire pipeline is outlined in Algorithm \ref{alg:inference_pipeline}. Given the set of input images $\{x_i\}_{i=1}^N$, we first pass each image $x_i$ through the encoder network $e_i = \enc(x_i)$ to obtain deep image embeddings $\{e_i\}_{i=1}^N$. For each image, we also compute the scene motion $m_i$ \wrt to first image $x_1$, and the certainty values $v_i$ used to weigh our feature space reconstruction loss (Sec.~\ref{sec:certainty_pred}). Next, we estimate the optimal latent encoding $\hat{z}$ of the output image $y$ which minimizes our reconstruction loss $L(z)$. This is achieved by using the iterative steepest-descent algorithm (Sec.~\ref{sec:optimization}). First, we obtain an initial latent encoding $z^0$ using an initializer network $P$. The initial encoding is then refined by applying $K_\text{SD}$ steepest-descent steps (Equation~\ref{eq:sd}) to obtain $z^{K_\text{SD}}$. The latent encoding $z^{K_\text{SD}}$ is finally passed through the decoder network $\hat{y} = \dec(K_\text{SD})$ to obtain the prediction $\hat{y}$. Note that each step in our inference pipeline is differentiable \wrt the parameters of the encoder $\enc$, certainty predictor $W$, initializer $P$, feature degradation $G$, and decoder $\dec$ networks. This allows us to learn each of these modules end-to-end from data, as described in Sec.~\ref{sec:training}.

\algrenewcommand{\algorithmicrequire}{\textbf{Input:}}
\newcommand{\assign}{\leftarrow}
\newcommand{\algcomment}[2]{\hspace{#2mm}{\footnotesize\# \textit{#1}}}
\begin{algorithm}[t]
	\caption{Inference pipeline}
	\begin{algorithmic}[1]
		\Require Multiple images $\{x_i\}_{i=1}^N$, Number of steepest-descent iterations $K_\text{SD}$
		\For{$i = 1, \ldots, N$}       
		\State $e_i \assign \enc(x_i)$ \algcomment{Map each input image to embedding space}{76}
		\State $m_i \assign \texttt{MotionEstimator}(x_i, x_1)$  \algcomment{Estimate scene motion for each image \wrt the first image}{40.5}
		\State $v_i \assign W(e_i, e_1, m_i)$  \algcomment{Estimate the certainty values for each image}{65.0}
		\EndFor
	    \\
		\State $z^0 \assign \modelinit(e_1))$ \algcomment{Obtain initial latent encoding}{79.5}
		\For{$j = 0, \ldots, K_\text{SD} - 1$}       \algcomment{For every steepest-descent iteration}{60}
		\State $g^j \assign -2\sum_{i=1}^N \shift_{\motion_i}^{\text{T}} G \conv^{\text{T}} \left(v_i^2 \!\cdot\! \left(e_i - G \conv \shift_{\motion_i}\!(z^j)\right)\right) + 2\lambda z^j$ \algcomment{Obtain gradient of loss \emph{(7)} in main paper \wrt z}{9.3}
		\State $\alpha^j \assign \frac{\|g^j\|_2^2}{\sum_{i=1}^N  2\|v_i \cdot \left(G \conv \shift_{\motion_i}\left(g^i\right)\right) \|_2^2 + 2\lambda \|g^j\|_2^2}$ \algcomment{Calculate optimal step-length along gradient direction $g^j$}{37.5}
		\State $z^{j+1} \assign z^j - \alpha^j g^j$ \algcomment{Update latent encoding using the estimated step-length $\alpha^j$}{66.0}
		\EndFor
		\State $\pred \assign \dec(z^{K_\text{SD}})$ \algcomment{Decode latent encoding to obtain the output image}{78}
	\end{algorithmic}
	\label{alg:inference_pipeline}
\end{algorithm}
\section{Network Architecture}
\label{sec:arch_details}
\subsection{RAW Burst Super-Resolution}
Here, we provide more details about the network architecture employed for the burst super-resolution task in Section \ref{sec:exp_sr} and \ref{sec:exp_abl}.

\parsection{Encoder $\enc$} The encoder packs each 2 $\times$ 2 block in the input RAW image along the channel dimension to obtain a 4 channel input. This ensures translation invariance, while also reducing the memory usage. The packed input is passed through a convolution+ReLU block to obtain a 64 dimensional feature map. This feature map is processed by 9 Residual blocks~\cite{He2016DeepRL}, before being passed through a final convolution+ReLU block to obtain a 256 dimensional embedding $\enc(x_i)$ of the input $x_i$. 

\parsection{Operator $G$} We use a convolution layer with stride 2 as our feature degradation operator $G$. The operator takes a $64$ dimensional embedding as input to generate a $256$ dimensional output, which is compared with the embedding $\enc(x_i)$ of the input image to compute the feature space reconstruction error.

\parsection{Initializer $P$} We use the sub-pixel convolution layer~\cite{Shi2016RealTimeSI} to generate the initial latent encoding $z^0$ of the output image $\pred$. The initializer takes the embedding $\enc(x_1)$ of the first burst image $x_1$ as input and upsamples it by a factor of 2 via sub-pixel convolution to output $z^0$.

\parsection{Certainty Predictor $W$} The certainty predictor takes the embedding of the input images $\{\enc(x_j)\}_{j=1}^N$, along with the motion estimation $m_i$ as input. Each image embedding $\enc(x_i)$ is passed through a convolution+ReLU block to obtain the 64-dimensional feature map $e_i$. The features $e_1$ extracted from the reference image $x_1$ are then warped to $i$-th image to compute the residual $e_i - \shift_{\motion_i}(e_1)$, which indicates possible alignment errors.
Additionally, we pass the modulo 1 of scene motion $m_i$ mod $1$ through a convolution+ReLU block followed by a residual block to obtain the 64-dimensional motion features $\tilde{m}_i$. The encoding $e_i$, the residual $e_i - \shift_{\motion_i}(e_1)$, and the motion feature $\tilde{m}_i$ are then concatenated along the channel dimension and projected to $128$ dimensional feature map using a convolution+ReLU block. The resulting feature map is passed through 3 residual blocks, followed by a final convolution layer to obtain the output $\tilde{v_i}$. Certainty value $v_i$ for the $i$-th image is then obtained as the absolute value $|\tilde{v_i}|$ to ensure positive certainties. 

\parsection{Decoder $\dec$} The decoder passes the encoding $\hat{\predenc}$ of the output image $\pred$ through a convolution+ReLU block to obtain a 64-dimensional feature map. This feature map is passed through 5 residual blocks, followed by an upsampling by a factor of $s / \tilde{s} = 4$ using the sub-pixel convolution layer~\cite{Shi2016RealTimeSI}. The upsampled 32-dimensional feature map is then passed through 5 additional residual blocks, followed by a final convolution layer which predicts the output RGB image $\pred$.

\subsection{Burst Denoising}
Here, we provide more details about the network architecture employed for the burst denoising task in Section \ref{sec:exp_denoise} and \ref{sec:exp_abl} in the main paper.

\parsection{Encoder $\enc$} The encoder concatenates the input image $x_i$ and the per-pixel noise estimate $n_i$ along the channel dimension, and passes it through a convolution+ReLU block. The resulting 32-dimensional output is processed by 4 residual blocks, followed by a convolution+ReLU block to obtain a 64-dimensional encoding $\enc(x_i, n_i)$ of the input image $x_i$.

\parsection{Operator $G$} We use a convolution layer as our feature degradation operator $G$. The operator takes a $16$ dimensional encoding $z$ of the output image $y$ as input to generate a $64$ dimensional output, which is compared with the embedding $\enc(x_i, n_i)$ of the input image to compute the feature space reconstruction error.

\parsection{Initializer $P$} We pass the embedding $\enc(x_1)$ of the first burst image $x_1$ through a convolution layer to obtain the initial output image encoding $z^0$.

\parsection{Certainty Predictor $W$} The certainty predictor takes the embedding of the input images $\{\enc(x_j, n_j)\}_{j=1}^N$, the noise estimate $n_i$, along with the motion estimation $m_i$ as input. Each image embedding $\enc(x_i, n_i)$ is passed through a convolution+ReLU block to obtain the 16-dimensional feature map $e_i$. The features $e_1$ extracted from the reference image $x_1$ are then warped to $i$-th image to compute the residual $e_i - \shift_{\motion_i}(e_1)$, which indicates possible alignment errors.
In parallel, the per-pixel noise estimate $n_i$ is passed through a convolution+ReLU block, followed by a residual block to obtain 32-dimensional noise features $\tilde{n_i}$.
Additionally, we pass the modulo 1 of scene motion $m_i$ mod $1$ through a convolution+ReLU block to obtain the 8-dimensional motion features $\tilde{m}_i$. 
The encoding $e_i$, the residual $e_i - \shift_{\motion_i}(e_1)$, the noise features $\tilde{n_i}$, and the motion features $\tilde{m}_i$ are then concatenated along the channel dimension and projected to $32$ dimensional feature map using a convolution+ReLU block. The resulting feature map is passed through 1 residual blocks, followed by a final convolution layer to obtain the output $\tilde{v_i}$. Certainty value $v_i$ for the $i$-th image is then obtained as the absolute value $|\tilde{v_i}|$ to ensure positive certainties. 

\parsection{Decoder $\dec$} The decoder passes the encoding $\hat{\predenc}$ of the output image $\pred$ through a convolution+ReLU block to obtain a 64-dimensional feature map. This feature map is passed through 9 residual blocks, followed by a final convolution layer which predicts the denoised image $\pred$.

\parsection{Motion Estimation in Ours$^\dag$} In Section \ref{sec:exp_denoise}, we report results for a variant of our approach \textbf{Ours$^\dag$} which employs a custom optical flow network to find the relative motion between image $x_i$ and the reference image $x_1$. We use a pyramidal approach with cost volume, commonly employed in state-of-the-art optical flow networks~\cite{Sun2018PWCNetCF,Ilg2017FlowNet2E}. The architecture of our optical flow network is described here. We first pass both $x_i$ and $x_1$ through a convolution+ReLU block to obtain 32-dimensional feature maps. These are then passed through 6 residual blocks. Before each of the first two residual blocks, we downsample the input feature maps by a factor of 2 using a convolution+ReLU block with stride 2 for computational efficiency. Next, we construct a feature pyramid with 2 scales, which is used to compute the optical flow. We pass the output of the last residual block through a convolution+ReLU block to obtain 64-dimensional feature maps $f^1_i$ and $f^1_1$. These feature maps are then passed through another convolution+ReLU block with stride 2 to obtain lower resolution feature maps $f^2_i$ and $f^2_1$. Next, we construct a partial cost volume containing pairwise matching scores between pixels in $f^2_i$ and $f^2_1$ using the correlation layer. For efficiency, we only compute matching scores of a pixel in $f^2_i$ with spatially nearby pixels in $f^2_1$ within a $7 \times 7$ window. 
The cost volume is concatenated with the feature map $f^2_i$ and passed through two convolution+ReLU blocks with 128 and 64 output dimensions. The output feature map is passed through a final convolution layer to obtain a coarse optical flow $m^2_i$. This initial estimate is upsampled by a factor of 2 and used to warp the feature map $f^1_1$ to the $i$-th image. We then compute the matching scores between $f^1_i$ and $\shift_{m^2_i}(f^1_1)$ using a $5 \times 5$ spatial window. A refined optical flow $m^1_i$ is then obtained using the same architecture as employed for pyramid level 2, without weight-sharing. The estimate $m^1_i$ is then upsampled by a factor of $4$ to obtain the final motion estimate $m_i$.

\section{Analysis of Certainty Predictor}
\label{sec:certainty_pred_supp}
\begin{figure*}[t]
    \centering%
    \subfloat[Input Image]{
	\resizebox{0.33\columnwidth}{!}{%
		\includegraphics[width=\textwidth]{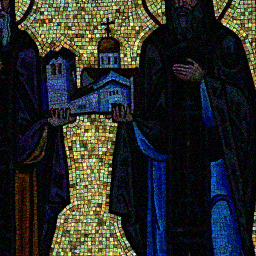}
	}\label{fig:certainty_input}%
	}
    \subfloat[Certainty Values, Original $m_i$]{
	\resizebox{0.33\columnwidth}{!}{%
		\includegraphics[width=\textwidth]{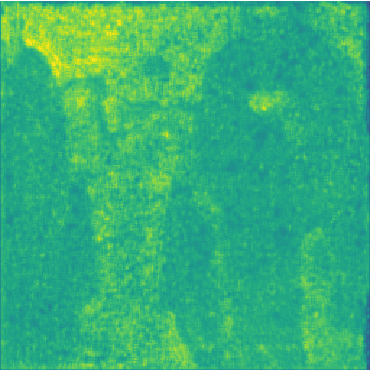}
	}\label{fig:certainty_normal}%
	}
    \subfloat[Certainty Values, Corrupted $m_i$]{
	\resizebox{0.33\columnwidth}{!}{%
		\includegraphics[width=\textwidth]{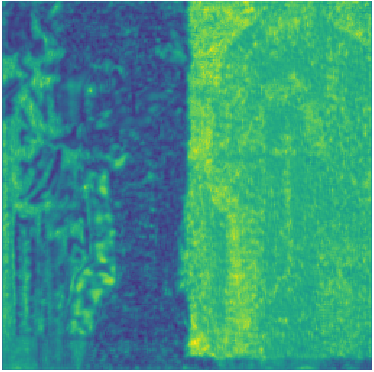}
	}\label{fig:certainty_corr}%
	}
    \caption{Analysis of our certainty predictor $W$. The channel-wise mean of the certainty values predicted by $W$ for the input image (a), using the estimated scene motion $m_i$ is shown in (b). Next, we corrupt the motion estimate $m_i$ for the left half of the image by adding a fixed offset of 16 pixels in both directions. The mean certainty values predicted using the corrupted motion estimate $m_i$ is shown in (c). Our certainty predictor can detect the errors in motion estimate $m_i$ and assign lower certainty values to the corresponding image regions.}\vspace{0mm}
    \label{fig:certainty}
\end{figure*}

In this section, we analyse the behaviour of our certainty predictor $W$. The certainty predictor computes the certainty values $v_i$ for each element in our residual $\enc(x_i) - G(\shift_{\motion_i}(z))$. This allows us to reduce the impact of \eg errors in motion estimate $\motion_i$, by assigning a lower weight for such regions in our MAP objective \eqref{eq:ours_loss_w}. We analyse the behaviour of $W$ by manually corrupting the motion estimate $\motion_i$. Figure~\ref{fig:certainty_normal} shows the channel-wise mean of the predicted certainty values $v_i$ for an input image (Figure ~\ref{fig:certainty_input}), using the estimated scene motion $m_i$. We observe that the mean certainty values are approximately uniform over the image, with some slight variations according to the image intensity values. Next, we corrupt the motion estimate $m_i$ for the left half of the image by adding a fixed offset of 16 pixels in both directions. The certainty values predicted using these corrupted motion $m_i$ is shown in Figure~\ref{fig:certainty_corr}. As desired, our certainty predictor detects the alignment errors and assign a lower certainty values to the corresponding image regions (left half of the image).

\section{Detailed Ablative Study}
\label{sec:ablation_supp}
\begin{table}[!t]
	\centering\vspace{-1mm}
	\subfloat[SyntheticBurstSR]{
	\resizebox{0.365\columnwidth}{!}{%
		\begin{tabular}{ll|ccc}
\toprule
$P$& $K_\text{SD}$&PSNR$\uparrow$&LPIPS$\downarrow$&SSIM$\uparrow$\\\midrule
& 3 & 39.79 & 0.073 & 0.951\\
\checkmark& 0 & 36.29 & 0.123  & 0.912\\
\checkmark& 1 & 39.60 & 0.074 & 0.950\\
\checkmark& 2  & 39.75 & 0.074 & 0.951 \\
\checkmark& 3 & 39.82& 0.071 & 0.952 \\
\checkmark& 4 & 39.77& 0.071 & 0.951 \\
\checkmark& 5 & 39.64 & 0.072 & 0.950\\

\bottomrule
\end{tabular}

	}\label{tab:num_sd_sr}%
	}
	\subfloat[Grayscale Denoising]{\resizebox{0.6\columnwidth}{!}{%
		\begin{tabular}{ll|ccccc}
\toprule
$P$& $K_\text{SD}$ &Gain $\propto$ 1&Gain $\propto$ 2&Gain $\propto$ 4&Gain $\propto$ 8& Average\\\midrule
&3 & 39.29 & 36.42 & 33.33 &  28.86 & 34.47 \\
\checkmark&0 & 35.16 & 32.27 & 29.34& 25.81 & 30.65 \\
\checkmark &1 & 39.32 & 36.48 & 33.38 &  29.39 & 34.64 \\
\checkmark &2 & 39.36 & 36.49 & 33.37 &  29.38 & 34.65 \\
\checkmark &3 & 39.37 & 36.51 & 33.38 &  29.69 & 34.74 \\
\checkmark &4 & 39.33 & 36.46 & 33.37 &  29.11 & 34.57 \\
\checkmark &5 & 39.42 & 36.54 & 33.40 &  28.25 & 34.40 \\

\bottomrule
\end{tabular}
	}\label{tab:num_sd_denoise}%
	}
	\vspace{1mm}%
	\caption{Impact of initializer $P$ and number of steepest-descent iterations $K_\text{SD}$ on the SyntheticBurst super-resolution (a) and grayscale denoising (b) datasets.}
	\label{tab:num_sd}%
	\vspace{0mm}
\end{table}


In this section, we provide a detailed ablative study analysing the impact different components in our architecture. Our analysis is performed on the SyntheticBurst super-resolution dataset~\cite{Bhat2021DeepBS}, as well as the grayscale burst denoising dataset~\cite{Mildenhall2018BurstDW}.

\begin{figure}[t]
    \centering%
    \includegraphics[trim = 0 0 0 0, width=0.6\columnwidth]{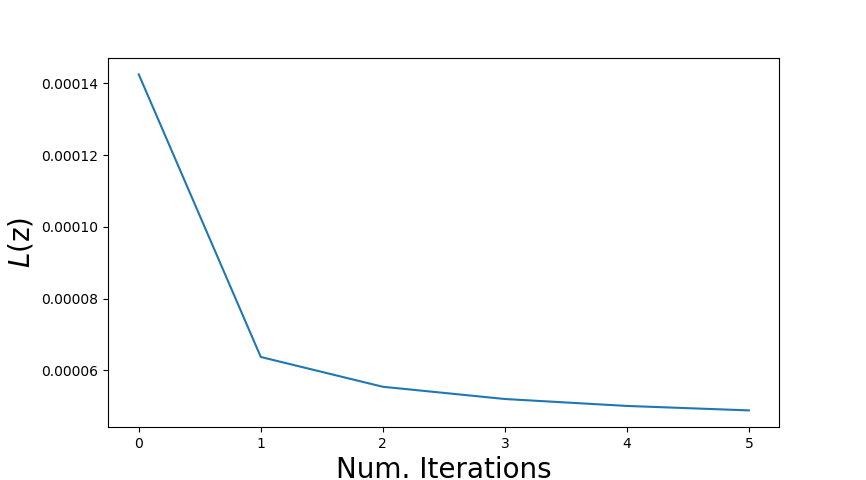}\vspace{0mm}
    \caption{Convergence analysis of the steepest-descent iterations. We plot our loss \eqref{eq:ours_loss_w}, \wrt the number of steepest-descent iterations. The loss is averaged over the $300$ burst sequences from the SyntheticBurst dataset.}\vspace{0mm}
    \label{fig:convergence}
\end{figure}

\parsection{Impact of number of iterations $K_\text{SD}$} We analyse the impact of the number of steepest-descent iterations $K_\text{SD}$ by training and evaluating our approach with different values of $K_\text{SD}$. The results on the SyntheticBurst super-resolution dataset~\cite{Bhat2021DeepBS} are provided in Table~\ref{tab:num_sd_sr}, while the results on the grayscale burst denoising dataset~\cite{Mildenhall2018BurstDW} is shown in Table~\ref{tab:num_sd_denoise}. Additionally, a convergence analyses of the steepest-descent steps is provided in Figure~\ref{fig:convergence}. The entries with $K_\text{SD} = 0$ directly output the initial encoding $z^0$ predicted by our initializer $P$ to the decoder. Since the initializer only utilizes the first burst image, the entry $K_\text{SD} = 0$ corresponds to the SingleImage baseline included in Table \ref{tab:synburst} and Table \ref{tab:kpn_grayscale}. Performing just a single steepest-descent step already provides a large improvement over the SingleImage baseline with a PSNR of 39.60 on the SyntheticBurst dataset. This demonstrates the fast convergence of the steepest-descent iterations. Both the super-resolution and denoising performance improves gradually with an increase in $K_\text{SD}$, and the best results are obtained with $K_\text{SD} = 3$ iterations. Performing more iterations $K_\text{SD} > 3$ results in a small decrease in performance, indicating that early-stopping can act as a regularizer, leading to improved performance. 

\parsection{Impact of initializer $P$} Here, we analyse the impact of our initializer module $P$, which computes the initial encoding $z^0$. We evaluate a variant of our approach which does not utilizes $P$, instead setting the initial encoding $z^0$ to zeros. The results on the SyntheticBurst dataset and the grayscale burst denoising dataset are shown in Table~\ref{tab:num_sd_sr} and Table~\ref{tab:num_sd_denoise}, respectively. Thanks to the fast convergence of the steepest-descent iterations, we observe that the initializer module $P$ only provides a small improvement in performance.

\begin{table}[!t]
	\centering\vspace{-1mm}
	\resizebox{0.36\columnwidth}{!}{%
		\begin{tabular}{lccc}
\toprule
$\tilde{s}$&PSNR$\uparrow$&LPIPS$\downarrow$&SSIM$\uparrow$\\\midrule
1 & 39.38 & 0.077 & 0.948\\
2& 39.82& 0.071 & 0.952 \\
4 & 39.89& 0.071 & 0.953 \\

\bottomrule
\end{tabular}
	}\vspace{1mm}%
	\caption{Impact of downsampling factor $\tilde{s}$ of feature degradation $G$ on the SyntheticBurst dataset.}
	\label{tab:stride_sr}%
	\vspace{0mm}
\end{table}

\parsection{Impact of downsampling factor $\tilde{s}$ of feature degradation $G$} We analyse the impact of the downsampling factor $\tilde{s}$ of operator $G$ on the burst super-resolution task. Results for different values of $\tilde{s}$ on the SyntheticBurst dataset is provided in Table~\ref{tab:stride_sr}. We observe that using a higher downsampling factor $\tilde{s}$ in operator $G$ leads to better results. However the improvement when using a downsampling factor $\tilde{s} = 4$ compared to $\tilde{s} = 2$ is relatively small ($+0.07$dB). Hence, we use $\tilde{s} = 2$ in our final version to obtain better computational efficiency.

\section{Qualitative Results}
\label{sec:qual_supp}
\begin{figure*}[t]
    \centering%
    \includegraphics[trim = 0 0 0 0, width=0.99\textwidth]{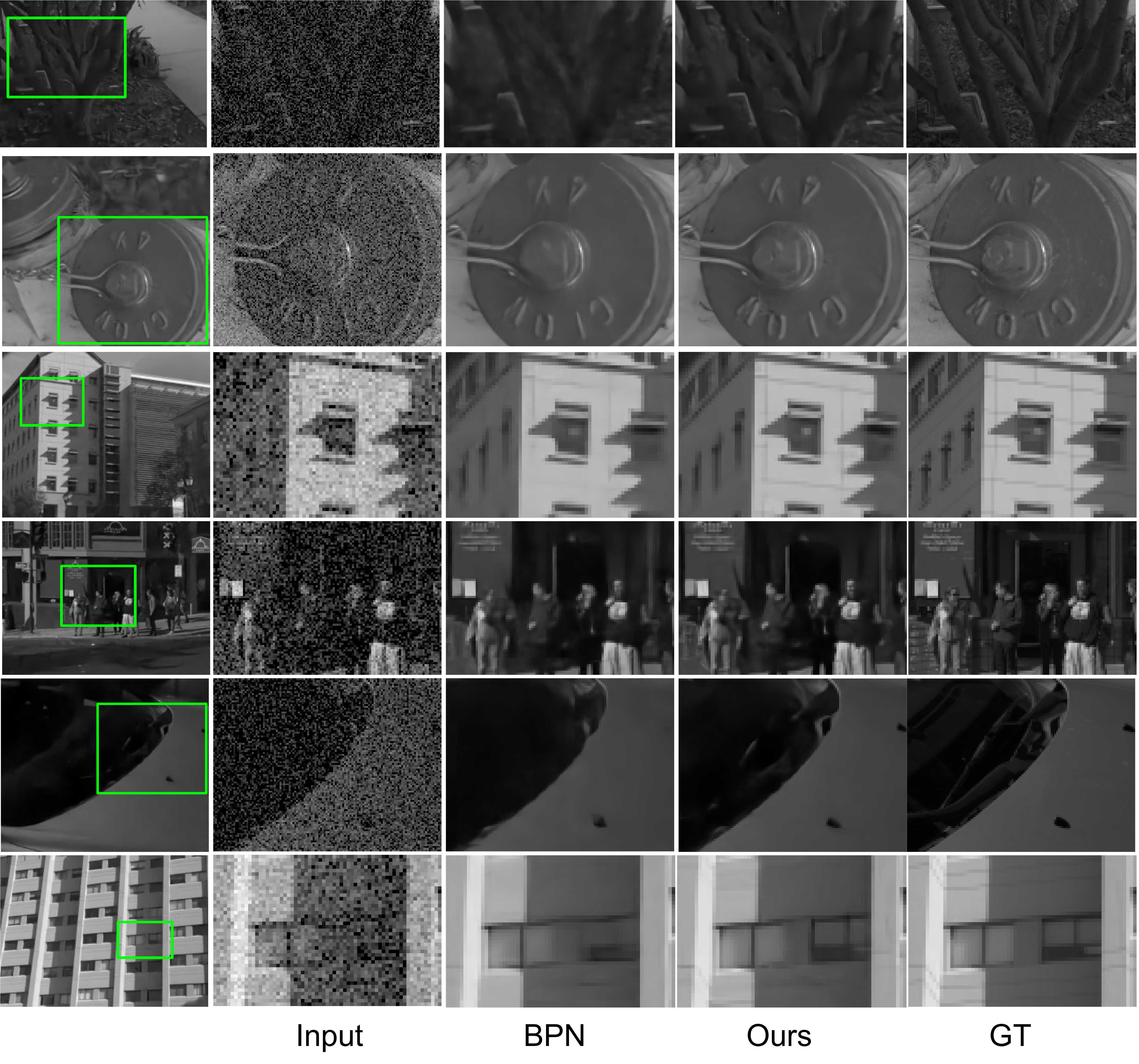}\vspace{0mm}
    \caption{Qualitative comparison of our approach with BPN~\cite{Xia2020BasisPN} on the grayscale burst denoising dataset~\cite{Mildenhall2018BurstDW}. Compared to BPN, our approach can recover higher frequency details, without oversmoothing the image.}\vspace{0mm}
    \label{fig:qual_gray}
\end{figure*}

\begin{figure*}[t]
    \centering%
    \includegraphics[trim = 0 0 0 0, width=0.99\textwidth]{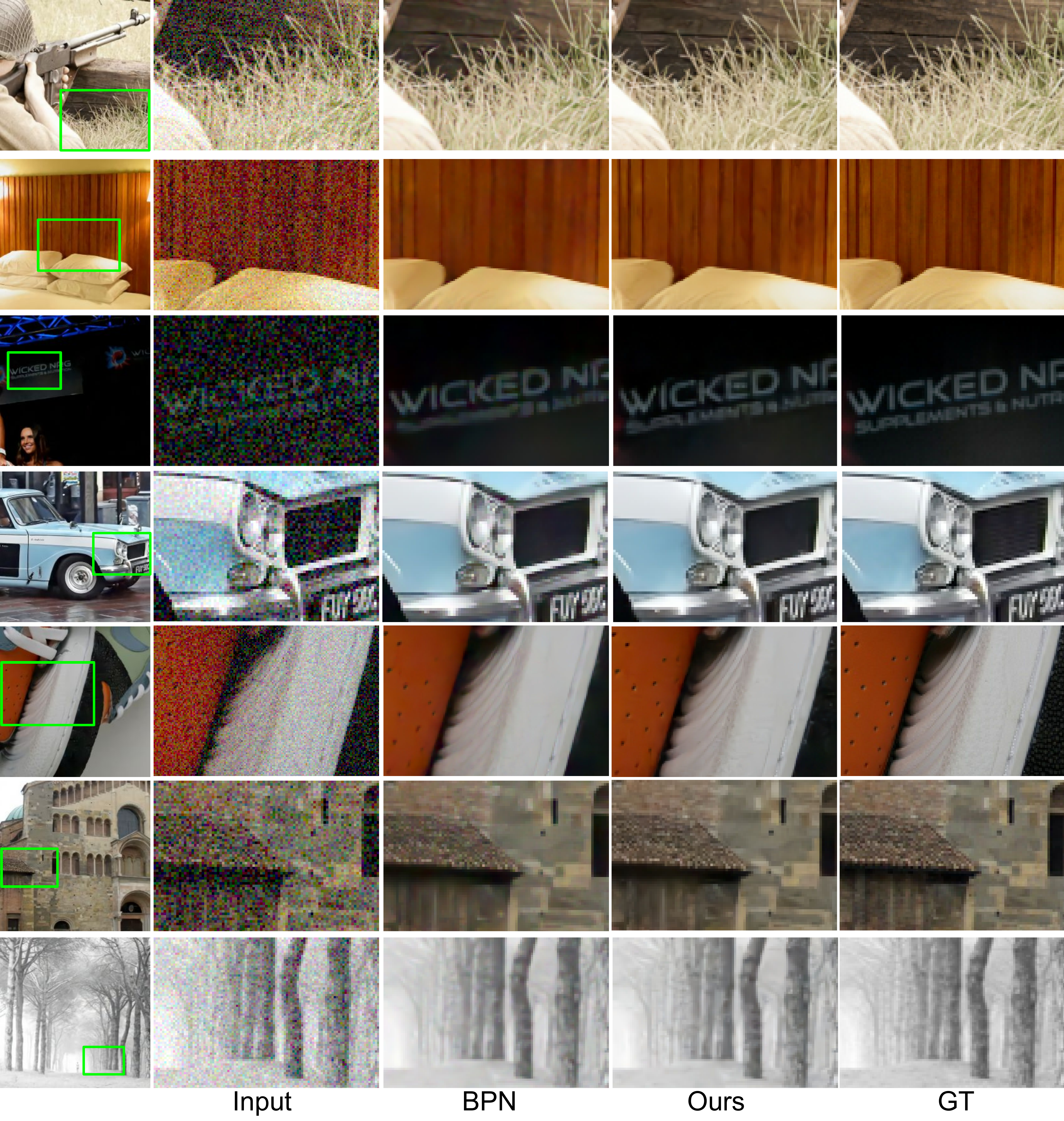}\vspace{0mm}
    \caption{Qualitative comparison of our approach with BPN~\cite{Xia2020BasisPN} on the color burst denoising dataset~\cite{Xia2020BasisPN}. Our approach can generate clean images without introducing any color artifacts.}\vspace{0mm}
    \label{fig:qual_color}
\end{figure*}

\begin{figure*}[t]
    \centering%
    \includegraphics[trim = 0 0 0 0, width=0.99\textwidth]{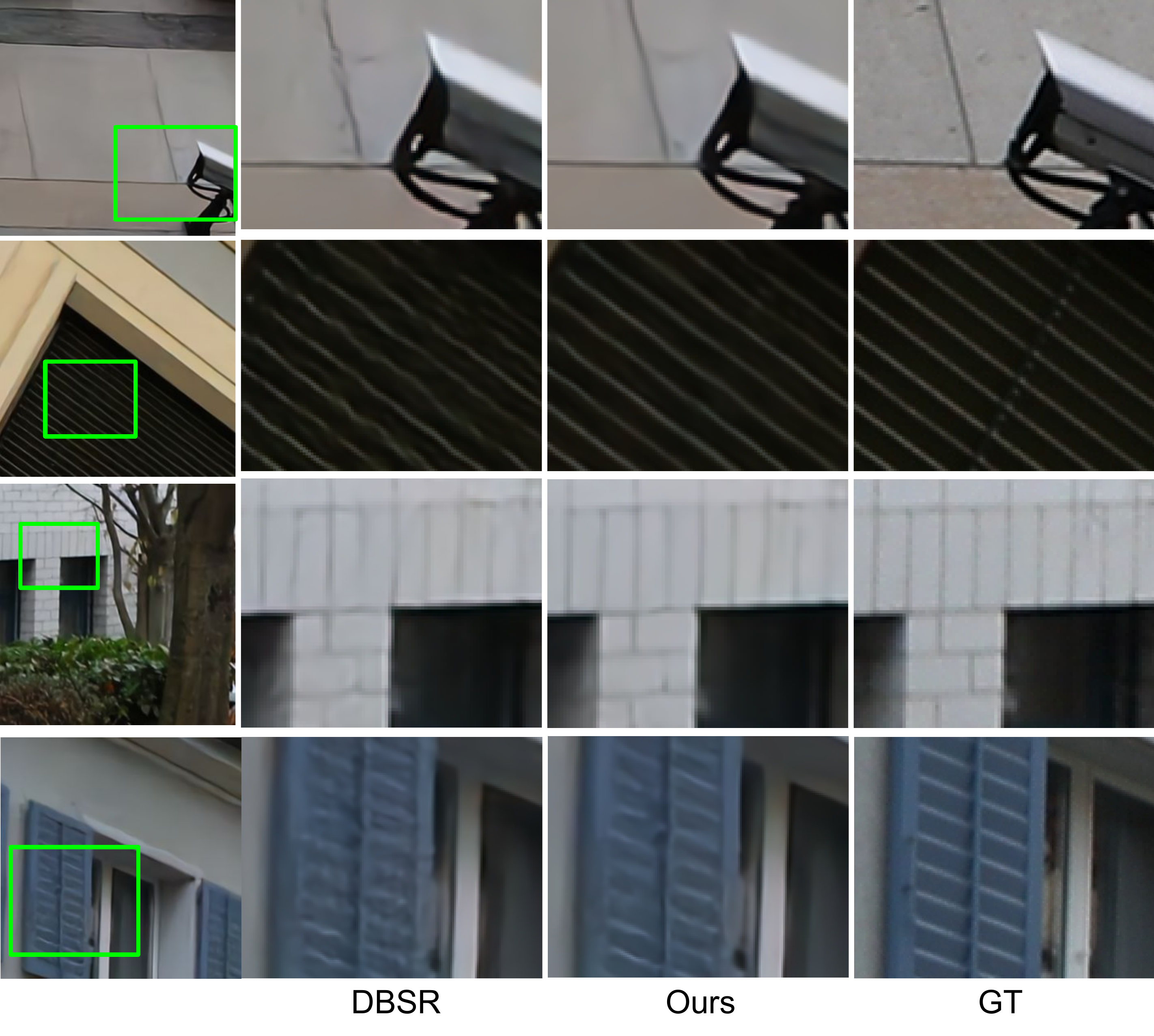}\vspace{0mm}
    \caption{Qualitative comparison of our approach with DBSR~\cite{Bhat2021DeepBS} on the SyntheticBurst super-resolution dataset~\cite{Bhat2021DeepBS}. Our approach can better recover high frequency details and generates sharper images.}\vspace{0mm}
    \label{fig:qual_syn_sr}
\end{figure*}

\begin{figure*}[t]
    \centering%
    \includegraphics[trim = 0 0 0 0, width=0.99\textwidth]{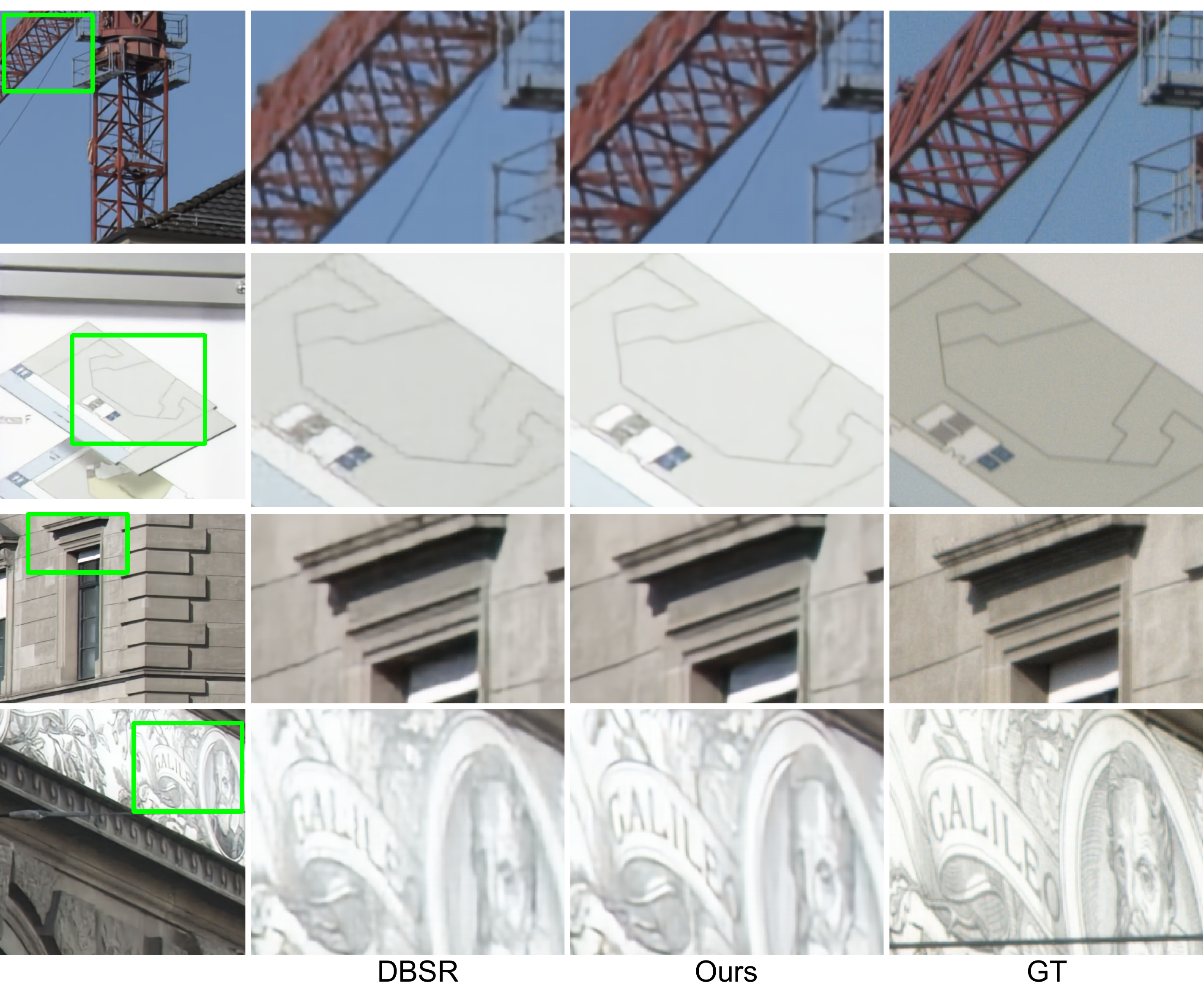}\vspace{0mm}
    \caption{Qualitative comparison of our approach with DBSR~\cite{Bhat2021DeepBS} on the real-workd BurstSR dataset~\cite{Bhat2021DeepBS}. Note that the ground truth image and the input burst are captured using different cameras, resulting in a color shift between the network predictions and the ground truth.}\vspace{0mm}
    \label{fig:qual_burstsr}
\end{figure*}

In this section, we provide additional qualitative results. A qualitative comparison with BPN~\cite{Xia2020BasisPN} on the grayscale~\cite{Mildenhall2018BurstDW} and color~\cite{Xia2020BasisPN} burst denoising datasets are provided in Figure~\ref{fig:qual_gray} and Figure~\ref{fig:qual_color}, respectively. Figure~\ref{fig:qual_syn_sr} contains a comparison of our approach to DBSR~\cite{Bhat2021DeepBS} on the SyntheticBurst RAW super-resolution dataset. Additional comparison on the real-world BurstSR dataset~\cite{Bhat2021DeepBS} is provided in Figure~\ref{fig:qual_burstsr}.

\end{document}